\pdfoutput=1
\documentclass[superscriptaddress,showkeys,nofootinbib,aps,prd]{revtex4-1}
\usepackage{amsmath}
\usepackage{amssymb}
\usepackage{graphicx}
\usepackage{color}
\usepackage{amsmath}
\usepackage{subcaption}

\newcommand{\braket}[1]{\left\langle#1\right\rangle}

\begin{document}
\title{\mathversion{bold}A chiral SU(4) explanation of the $b\to s$ anomalies}

\author{Shyam Balaji} 
\email{shyam.balaji@sydney.edu.au}
\affiliation{ARC Centre of Excellence for Particle Physics at the Terascale, School of Physics, The University of Sydney, NSW 2006, Australia}

\author{Robert Foot} 
\email{rfoot@unimelb.edu.au}
\affiliation{ARC Centre of Excellence for Particle Physics at the Terascale, School of Physics, The University of Sydney, NSW 2006, Australia}
\affiliation{ARC Centre of Excellence for Particle Physics at the Terascale, School of Physics, The University of Melbourne, VIC 3010, Australia}
\author{Michael A.~Schmidt} 
\email{m.schmidt@unsw.edu.au}
\affiliation{School of Physics, The University of New South Wales, Sydney, NSW 2052, Australia}

\begin{abstract}
	We propose a variant of the Pati-Salam model, with gauge group $SU(4)_C\times SU(2)_L\times U(1)_{Y'}$,
in which the chiral left-handed quarks and leptons are unified into a $\underline{4}$ of $SU(4)_C$,
while the right-handed quarks and leptons have quite a distinct treatment.
The  $SU(4)_C$ leptoquark gauge bosons can explain the measured deviation
of lepton flavour universality in the rare decays: $\bar B \to \bar K^{(*)} \bar \ell \ell$, $\ell = \mu, e$
(taken as a hint of new physics).
The model satisfies the relevant experimental constraints and makes predictions for the important $B$ and $\tau$ decays and results in a correlation between leptonic $B_s$ decays and $R_{K}$. 
These predictions 
will be tested  at the LHCb and Belle II experiments when increased statistics become available.
\end{abstract}


\maketitle

\section{Introduction}

There is mounting evidence for a violation of lepton flavour universality (LFU) in flavour-changing neutral 
current processes $b\to s\bar \mu\mu$ in recent measurements of $B$ decays~\cite{Aaij:2014ora,Aaij:2017vbb,Aaij:2013qta,Aaij:2015oid,Wehle:2016yoi,Aaij:2014pli,Aaij:2015esa}. 
The theoretically cleanest probes are the LFU ratios
\begin{equation}
    R_{K^{(*)}} = \frac{\Gamma(\bar B\to \bar K^{(*)} \mu^+\mu^-)}{\Gamma(\bar B\to \bar K^{(*)}e^+e^-)}
\end{equation}
which compare the decay rate $b\to s\bar \ell\ell$ ratio between muons and electrons respectively. 
Hadronic uncertainties cancel out in the ratios as long as new physics effects are small~\cite{Hiller:2003js,Capdevila:2016ivx,Capdevila:2017bsm}. 
The current experimental data shown in Table~\ref{tab:RK} indicates deviations of more than $2\sigma$ for both LFU 
ratios $R_{K^{(*)}}$ separately. An effective field theory analysis including all $b\to s\bar \ell\ell$ data in fact shows that the introduction of operators 
\begin{align}
	O_9 &= [\bar s \gamma^\mu P_L b] [\bar \mu\gamma_\mu \mu] &
	O_{10} &= [\bar s \gamma^\mu P_L b] [\bar \mu\gamma_\mu\gamma_5 \mu]
\end{align}
may improve the global fit by $4-5\sigma$ \cite{Altmannshofer:2017yso,DAmico:2017mtc,Hiller:2017bzc,Capdevila:2017bsm,Geng:2017svp,Ciuchini:2017mik}. 
	In addition to the $R_K$ anomaly, there is some evidence for a deviation from standard model (SM) predictions in the muon $g-2$ measurements (see e.g.~Ref.~\cite{Blum:2013xva}) and also in
charged-current semi-leptonic decays $b\to c\ell\bar \nu$ ($R_D$ anomaly)  see e.g.~Ref.~\cite{Bifani:2018zmi}.
The leading SM contributions to $b \to c \ell \bar \nu$ arise at tree level, while the contributions to the muon $g-2$ and $b\to s\bar\ell\ell$ arise at one-loop level. Although new physics contributions to the muon $g-2$ arise at loop level, there may be new physics contributions to $b\to c\ell\bar\nu$ and $b\to s\bar\ell \ell$ at tree level.
It follows that the $b \to s$ processes are expected to provide a more sensitive probe of deviations from the SM.
The experimental sensitivity is expected to significantly improve in the next few years: LHCb will acquire 
more data and the Belle II experiment is anticipated to start collecting data with the full detector soon and will measure $R_{K^{(*)}}$ with a precision of $3.6\%$ ($3.2\%$). 
\begin{table}[b!]\centering
	\begin{ruledtabular}
		\begin{tabular}{cccc}
		&  observed & SM & $q^2$ range\\ \hline
		$R_{K}$ & $0.745 ^{+0.090} _{-0.074}\pm0.036$ \cite{Aaij:2014ora}	&	$1.0003\pm 0.0001$ \cite{Bobeth:2007dw} & $1\, \mathrm{GeV}^2 < q^2<6\, \mathrm{GeV}^2$  \\
		$R_{K^*}$ 
		&$0.69 ^{+0.11} _{-0.07}\pm 0.05$ \cite{Aaij:2017vbb}
		& $1.00 \pm 0.01$ \cite{Bordone:2016gaq}
		& $1.1\, \mathrm{GeV}^2 < q^2 < 6\, \mathrm{GeV}^2$ 
		\\
\end{tabular}
\end{ruledtabular}
\caption{LFU ratios $R_{K^{(*)}}$, where we first list the statistical error and then the systematic.}
\label{tab:RK}
\end{table}

The possibility that some or even all of these deviations might be a harbinger of new physics has been 
entertained in the literature, e.g.~by introducing a new effective interaction of third-generation weak eigenstates~\cite{Glashow:2014iga}, 
models of $Z^\prime$ gauge bosons e.g. ~\cite{Descotes-Genon:2013wba,Gauld:2013qba,Chiang:2016qov} 
and leptoquarks e.g.~\cite{Datta:2013kja,Hiller:2014yaa}.
In this paper we consider
a rather particular kind of Pati-Salam inspired $SU(4)$ gauge model, with chiral gauge interactions with quarks and leptons. 
In this scheme, the $b \to s$ anomaly is explained via tree level leptoquark 
gauge bosons with mass $m_{W'} \gtrsim 10$ TeV.
Although 
various kinds of $SU(4)$ models have also been considered in the context of the B-physics anomalies in several 
papers~\cite{Assad:2017iib,DiLuzio:2017vat,Calibbi:2017qbu,Bordone:2017bld,Barbieri:2017tuq,Blanke:2018sro,Greljo:2018tuh,Bordone:2018nbg,Faber:2018qon,Heeck:2018ntp},
the proposal identified in this paper appears to have escaped attention in the literature.
Our model provides a very simple and predictive scheme, describing the $b \to s$ anomaly 
with only two parameters, $m_{W'}$ and a CKM-type mixing angle, $\theta$. The leptoquark gauge boson 
does not contribute significantly to the  $R_D$ anomaly. If both $R_D$ and $R_K$ anomalies are confirmed 
then the $R_K$ anomaly could be explained in terms of chiral Pati-Salam gauge bosons as described here, 
with $R_D$ explained, potentially, via scalar leptoquarks incorporated in simple extensions of the proposed
model.

The paper is organised as follows. In Sec.~\ref{sec:model} we introduce the model and discuss the relevant effective operators 
in Sec.~\ref{sec:EffectiveOperators}. Our results are presented in Sec.~\ref{sec:results} and we conclude in Sec.~\ref{sec:conclusions}.


\section{The Model}
\label{sec:model}

The Pati-Salam model~\cite{Pati:1974yy} is a left-right symmetric model based on the gauge group 
$SU(4)_C\times SU(2)_L\times SU(2)_R$ where
both chiral left- and right-handed leptons are interpreted as the fourth colour of $(4,2,1), (4,1,2)$ fermion multiplets
(the other three colours representing the quarks).
In the original version of the model,
quite stringent limits on the $SU(4)$ symmetry breaking scale arises from various
processes, especially two-body leptonic decays of mesons: $K \to \bar \mu e$, $B \to \bar \mu e$ etc..
These two-body
rare decays are effectively enhanced over three-body processes because the $SU(4)$ leptoquark gauge bosons couple in a 
vector-like manner to the charged leptons, eliminating any helicity suppression.

It was noticed some time ago~\cite{Foot:1997pb,Foot:1999wv} that variants
of the Pati-Salam model can easily be constructed whereby the $SU(4)$ leptoquark gauge bosons couple in a chiral
fashion to the quarks and leptons. Such chiral $SU(4)_C$ models are less constrained than the original Pati-Salam
model, and $SU(4)$ symmetry breaking at the TeV scale can be envisaged.
The particular model studied in Refs.~\cite{Foot:1997pb,Foot:1999wv}
featured leptoquark gauge bosons coupling to chiral right-handed quarks and leptons, a circumstance which is not well
suited to explaining the $R_K$ anomaly.
Here we aim to construct the simplest chiral $SU(4)$ model in which the leptoquark gauge bosons couple to
quarks and leptons in a predominately left-handed manner.

The gauge symmetry of the model is 
$SU(4)_C\times SU(2)_L\times U(1)_{Y'}$, and the
fermion/scalar particle content is listed in Table~\ref{tab:particles}.
\begin{table}
        \centering
	\begin{ruledtabular}
        \begin{tabular}{cc|cc}
		fermion & $(SU(4)_C, SU(2)_L, U(1)_{Y'})$
			& scalar & $(SU(4)_C,SU(2)_L,U(1)_{Y'}$)
                \\\hline
                ${\bf Q}_L$ & $(4,2,0)$ &
                $\phi$ & $(1,2,1)$\\
                ${\bf u}_R$ & $(4,1,1)$ &
                $\chi$ & $(4,1,1)$\\
                ${\bf d}_R$ & $(4,1,-1)$ &
                $\Delta$ & $(4,2,2)$\\
                $E_L$ & $(1,1,-2)$\\
                $e_R$ & $(1,1,-2)$\\
                $N_L$ & $(1,1,0)$\\
        \end{tabular}
\end{ruledtabular}
\caption{Particle content}
        \label{tab:particles}
\end{table}
The $SU(4)$ symmetry
is broken by the vacuum expectation value (VEV) of the scalar $\chi$ at a high scale ($\langle \chi\rangle \equiv w \gtrsim 10$ TeV), while the electroweak symmetry is broken by
the VEVs of the scalars $\phi$ and $\Delta$, with $\sqrt{v^2 + u^2} \simeq 174 \ {\rm GeV}$ where
$\langle \phi \rangle \equiv v$ and  $\langle \Delta \rangle \equiv u$.\footnote{The VEV $u$ also breaks $SU(4)_C\times U(1)_{Y^\prime}$, but its effects are suppressed, since we assume $u\ll w$.}
The symmetry breaking pattern that results is
\begin{equation}
\begin{gathered}
    SU(4)_C \times SU(2)_L \times U(1)_{Y'}  \\
   \downarrow \braket{\chi}\\
   SU(3) \times SU(2)_L \times U(1)_Y \\
   \downarrow \braket{\phi}, \braket{\Delta}\\
   SU(3) \times  U(1)_Q \\
\end{gathered}
\end{equation}
Here hypercharge $Y=T+Y'$ and electric charge $Q=I_3+\frac{Y}{2}$.
If we use the gauge symmetry to rotate the VEV of $\chi$ to the fourth component, then $T$ is the diagonal traceless $SU(4)$ generator with elements $(\frac{1}{3},\frac{1}{3},\frac{1}{3},-1)$.

The Yukawa Lagrangian is
\begin{align}
        \mathcal{L} &=
        Y_u \bar {\bf Q}_L  \tilde \phi {\bf u}_R
        + Y_d \bar {\bf Q}_L \phi {\bf d}_R
        + Y_N \bar {\bf u}_R \chi N_L
        + Y_E \bar {\bf d}_R \chi E_L
        + Y_e \bar {\bf Q}_L \Delta e_R
        \nonumber \\
&
        + m_1 \bar E_L e_R
        + \frac12 m_N \bar N_L^c N_L
        +h.c.\;,
\end{align}
where $\tilde\phi\equiv i\tau_2 \phi^*$, and we have used bold face notation to label $SU(4)_C$ $\underline{4}$  multiplets
which contain the usual quarks plus a leptonic component.
The generation index has been suppressed, and it is implicit that each of these components comes in three generations, i.e. $u_R \equiv u_R^i = (u_R, c_R, t_R)$,
$d_R \equiv d_R^i = (d_R, s_R, b_R)$, etc..
The $\chi$ field gives mass to the charged ($\frac{2}{3} e$) $W'$ and neutral $Z'$ gauge bosons along with the exotic charged $E_{L,R}^{-}$ and neutral $N_{L,R}$ fermions.
The SM fields acquire mass via the $\phi$ and $\Delta$ fields.

The quark mass matrices are given by $m_u= Y_u v$ and $m_d=Y_d v$,
while the charged and neutral lepton mass matrices are
\begin{align}
        M_{e,E} &= \begin{pmatrix}
                 Y_e u & m_d\\
                m_1 & Y_E^\dagger w    \\
        \end{pmatrix} &
        M_N &= \begin{pmatrix}
                0 & m_u & 0 \\
                m_u^T & 0 & Y_N w \\
                0 & Y_N^T w & m_N
        \end{pmatrix}
\ .
\end{align}
In defining these matrices we have adopted a
basis $(e, E)_{L,R}$ and $(\nu_L,N_R^c,N_L)$ where $e_L, \nu_L$ are the fourth components of ${\bf Q}_L$ and
$E_R, N_R$ are the fourth components of ${\bf d}_R$, ${\bf u}_R$. 
In the limit $w \gg m_1, m_d$ (assumed in this paper) the charged lepton masses reduce to $m_{e} \simeq Y_e u$, while the exotic charged leptons have mass $M_E \simeq Y_E^\dagger w$. 
Also, the $W'$ leptoquark $SU(4)$ gauge bosons couple chirally to the SM quarks and leptons.
It is beneficial to explicitly write out the fermion
multiplets. For the first generation we have
\begin{align}
        {\bf Q}_L &= \begin{pmatrix}
                u_r&d_r \\
                u_g&d_g\\
                u_b&d_b \\
                \nu&e\\
        \end{pmatrix}_{\!\!L} &
        {\bf d}_R &= \begin{pmatrix}
                d_r\\
                d_g\\
                d_b\\
                E\\
        \end{pmatrix}_{\!\!R}&
        {\bf u}_R &= \begin{pmatrix}
                u_r\\
                u_g\\
                u_b\\
                N\\
        \end{pmatrix}_{\!\!R}
	&&E_L
	&&e_R
	&&N_L
\ .
\end{align}
Note that the active neutrino masses are generated via an inverse seesaw, and their observed sub-eV mass scale
is compatible with a TeV scale VEV $w$. 

In this model the masses of the charged leptons
arise from the VEV of the $\Delta$ scalar, while the masses of the quarks result from the VEV of $\phi$.
In such a situation, consistent Higgs phenomenology requires the existence of a decoupling limit
where the LHC Higgs-like scalar is identified with the lightest neutral scalar in the model.
To see how this can arise consider the Higgs potential terms
\begin{eqnarray}
V(\chi,\phi,\Delta) = \lambda_1 (\chi^{\dagger}\chi - w^2)^2 + \lambda_2(\phi^{\dagger}\phi - v^2)^2 + m_{\Delta}^2 \Delta^{\dagger} \Delta 
- m_{123} \Delta^{\dagger} \phi \chi
- m_{123}^{*} \chi^{\dagger} \phi^{\dagger} \Delta
\ .
\end{eqnarray}
Here $m_{123}$ is a trilinear coupling of dimensions of mass which, without loss of generality, we can take to be real. For $\lambda_1, \lambda_2, m_{\Delta} > 0$,
and considering initially $m_{123} = 0$, the potential is minimised when $\langle \chi^{\dagger}\chi \rangle = w^2$, $\langle \phi^{\dagger}\phi \rangle = v^2$,
and $\langle \Delta \rangle = 0$. Taking advantage of the gauge symmetry, the VEVs can be rotated into the real part of one of the complex components of $\chi$ and $\phi$:
$\langle Re: \chi_0 \rangle = w$, $\langle Re: \phi_0 \rangle = v$.  In the non-trivial case where $m_{123} \neq 0$, a VEV is induced for the real part of $\Delta_0$
\begin{eqnarray}
\langle Re: \Delta_0 \rangle \equiv u \simeq \frac{m_{123}wv
}{m_{\Delta}^2}
\ .
\end{eqnarray}
In such a manner, $u \ll v$ can naturally arise if $m_{123}w/m^2_{\Delta}  \ll 1$.

The physical scalar content consists of electrically charged $5/3$ and $2/3$ coloured leptoquark scalars, a singly charged scalar, $\Delta^+$, three neutral scalars,
$\tilde\chi_0/\sqrt{2} = Re: \chi_0$,  $\tilde\phi_0/\sqrt{2} = Re: \phi_0$,
$\tilde\Delta_0/\sqrt{2} = Re: \Delta_0$, and a pseudo scalar, $\tilde\Delta_0'/\sqrt{2} = Im: \Delta_0$.
In the limit $w^2 \gg v^2$, the $\tilde\chi_0$ scalar decouples and
the two remaining neutral scalars mix so that their physical mass eigenstates take the form
\begin{eqnarray}
h &=& \cos\beta \tilde\phi_0 + \sin\beta \tilde\Delta_0
\nonumber \\
H &=& -\sin\beta \tilde\phi_0 + \cos\beta \tilde\Delta_0
\
\end{eqnarray}
where $\sin\beta \simeq m_{123}w/(m_{\Delta}^2) = u/v$
in the decoupling limit $m_{\Delta}^2 \gg m_{123} w$.
In this limit it is easy to check that the lightest scalar, $h$, has Higgs-like coupling to the SM particles.
This result would hold for the most general Higgs potential so long as a decoupling regime as described is considered~\cite{Haber:1989xc}. The
scalar $h$ can thus be identified with the Higgs-like scalar discovered at the LHC~\cite{Aad:2012tfa,Chatrchyan:2012xdj}.

Finally, the model features an unbroken global $U(1)_B$ baryon number symmetry. As with the standard model, this global symmetry
is not imposed but appears as an accidental symmetry of the Lagrangian. However, unlike the standard model, the unbroken baryon global symmetry
does not commute with the gauge symmetries, and is generated by
\begin{eqnarray}
B = \frac{B' + T}{4}  \ .
\end{eqnarray}
Here, we have introduced the generator, $B'$, which commutes with the gauge symmetries, and is defined by the charges: 
$B'({\bf Q}_L,{\bf u}_R,{\bf d}_R, \chi,\Delta) = 1$, $B'(E_L,e_R,N_L,\phi, {\cal G}) = 0$
(${\cal G}$ is the set of gauge fields).
With $B$ defined as above, one can easily check that $U(1)_B$ is an unbroken symmetry of the Lagrangian
(i.e. $B\langle \chi \rangle = B\langle \Delta \rangle = B \langle \phi \rangle = 0$).
The $U(1)_{B'}$ is also a symmetry of the Lagrangian, but is not independent of the gauge symmetries and $U(1)_B$.


\section{Effective operators}
\label{sec:EffectiveOperators}
The relevant new physics contributions to the anomalies and possible constraints are most efficiently described by the effective Lagrangian
\begin{equation}
	\mathcal{L}_{eff} = \frac{4 G_F}{\sqrt{2}} \frac{\alpha_{em}}{4\pi} \sum_{q,q\prime,\ell, \ell^\prime} V_{tq} V_{tq^\prime}^*  \sum_{i=9,10}(
	C_i^{qq^\prime \ell \ell^\prime} O^{qq^\prime \ell\ell^\prime}_i
	+C_i^{\prime qq^\prime \ell \ell^\prime} O^{\prime qq^\prime \ell\ell^\prime}_i
	)+\mathrm{h.c.}\;,
\end{equation}
where $O_i$ denotes operators with two down-type quarks and two charged leptons
\begin{align}
	O_9^{qq^\prime\ell\ell^\prime} & = (\bar q\gamma_\mu P_L q^\prime ) (\bar \ell \gamma^\mu \ell^\prime) &
    O_9^{\prime qq^\prime\ell\ell^\prime} & = (\bar q\gamma_\mu P_R q^\prime ) (\bar \ell \gamma^\mu \ell^\prime) \nonumber \\
    O_{10}^{qq^\prime\ell\ell^\prime} & = (\bar q\gamma_\mu P_L q^\prime ) (\bar \ell \gamma^\mu \gamma_5\ell^\prime) &
    O_{10}^{\prime qq^\prime \ell\ell^\prime} & = (\bar q\gamma_\mu P_R q^\prime ) (\bar \ell \gamma^\mu \gamma_5\ell^\prime) 
\ .
\end{align}
In the above, $G_F$ denotes the Fermi constant,  $\alpha_{em} =1/127.9$ the fine-structure constant evaluated at the electroweak scale, $V_{ij}$ are CKM mixing matrix elements, $q^{(\prime)}$ are down-type quark fields, $\ell^{(\prime)}$ denotes charged leptons and $P_{L,R} = (1\pm \gamma_5)/2$
are the chiral projection operators. 

The relevant $SU(4)$ gauge interactions with the fermions, 
together with the leptoquark gauge boson mass term, are given by
\begin{align}
	\mathcal{L} = 
	\frac{g_s}{\sqrt{2}}K_{ij} W^\prime_\mu \bar d_i \gamma^\mu P_L \ell_j
	+ \frac{g_s}{\sqrt{2}} K^{*}_{ji} W^{\prime *}_\mu \bar \ell_i \gamma^\mu P_L d_j
	- m_{W'}^2 W^{\prime *}_\mu W^{\prime \mu} 
\end{align}
where $g_s$
is the $SU(4)$ gauge coupling constant.
Here we have defined $\ell$ to include the three charged SM leptons and the three heavy exotic charged lepton mass eigenstates, i.e. $\ell = e,E$.
This means that $K_{ij}$ is in general a $3\times 6$ matrix which satisfies the unitarity condition $K K^\dagger  = 1_{3\times 3}$, where
$1_{3\times 3}$ is the $3\times 3$ unit matrix.

In this model the Wilson coefficients for the effective four-fermion interaction after integrating out the heavy $W'$ mediator and using the 
appropriate Fierz rearrangement to collect quark and lepton bilinears are
	\begin{align}\label{eq: C9 and C10}
	C_9^{qq^\prime\ell \ell^\prime} & = - C_{10}^{qq^\prime\ell \ell^\prime} =
	\frac{\sqrt2 \pi^2 \alpha_s  }{ V_{tq} V_{tq^\prime}^* \alpha_{em}} \frac{K_{q\ell^\prime} K_{q^\prime\ell}^*}{G_F m_{W^\prime}^2 } 
\end{align}
where $\alpha_s = g_s^2(m_{W'}^2)/4\pi$.
Typically, limits from lepton flavour violating Kaon decays are more stringent then those from $B$ meson decays, and this constrains the
possible flavour structure of the theory.
In order to satisfy these constraints, and to explain the 
$R_{K^{(*)}}$ 
anomaly, a particular structure of the $K$ matrix is suggested. Considering only the first 
3 columns of the general $K$ matrix, i.e. the part relevant to quark-SM lepton interactions, we adopt the limiting case:
\begin{align}
	K &= \begin{pmatrix}
		0 & 0 & 1 \\
		\cos\theta & \sin\theta & 0 \\
		-\sin\theta & \cos\theta & 0 \\
	\end{pmatrix} \;.
\label{m5}
\end{align}
In general, the zero elements need not be exactly zero, but for the $m_{W'},\ \theta$ values of interest for the $R_{K^{(*)}}$ measurements are
constrained from lepton flavour violating Kaon decays to be relatively small ($\lesssim 0.1$).


\section{Results \&  discussion}
 
\label{sec:results}

With the ansatz Eq.~(\ref{m5}) it is straightforward to evaluate the $W'$ leptoquark gauge boson contributions
to the $R_{K^{(*)}}$ anomaly. The model has the distinctive feature
that both $b\to s\bar ee$ and $b\to s \bar\mu\mu$ processes receive corrections of approximately the same magnitude, but with  opposite sign.  
One consequence of this is that modifications to the angular distributions are anticipated in both muon and electron channels. However, it is noteworthy that the muon channel is experimentally advantageous over the electron channel due to improved resolution.

The favoured region of parameter space for the model is identified using the \texttt{flavio} package \cite{david_straub_2018_1326349} and tree-level analytical estimations where appropriate. 
The $\bar{B} \to {\bar K}^{(*)} \mu ^+ \mu ^-$, $\bar{B} \to {\bar K}^{(*)} e ^+ e ^-$ rates are used to determine the $R_K$ and $R_{K^{*}}$ ratios for 
a given $m_{W'}$ leptoquark mass and $\theta$ mixing angle, with the $C_9$ and $C_{10}$ coefficients detailed in Eq.~(\ref{eq: C9 and C10}). 
Additionally, we calculate $BR({B^+}\to {K^+}\mu^- e^+)$ and $BR({B^+}\to {K^+} e^- \mu^+)$ values.
The $1 \sigma$ and [90\% C.L.] favoured parameter region is defined by the $m_{W'}, \theta$ values which satisfy
$R_K = 0.745 \pm 0.097$ [$R_K = 0.745 \pm 0.159$], $R_{K^*} = 0.69 \pm 0.12$ [$R_{K^*} = 0.69 \pm 0.20$] 
and also satisfy the 
current  90\% C.L. experimental limits $BR({B^+}\to {K^+}\mu^- e^+)<1.3\times10^{-7}$ and $BR({B^+}\to {K^+} e^- \mu^+)<9.1\times10^{-8}$ \cite{Tanabashi:2018oca}.
It turns out that the favoured region, defined in the way we have done, is not currently constrained by any other process.
\begin{figure}[tpb!]
    \centering
    \includegraphics[width=0.7\linewidth] {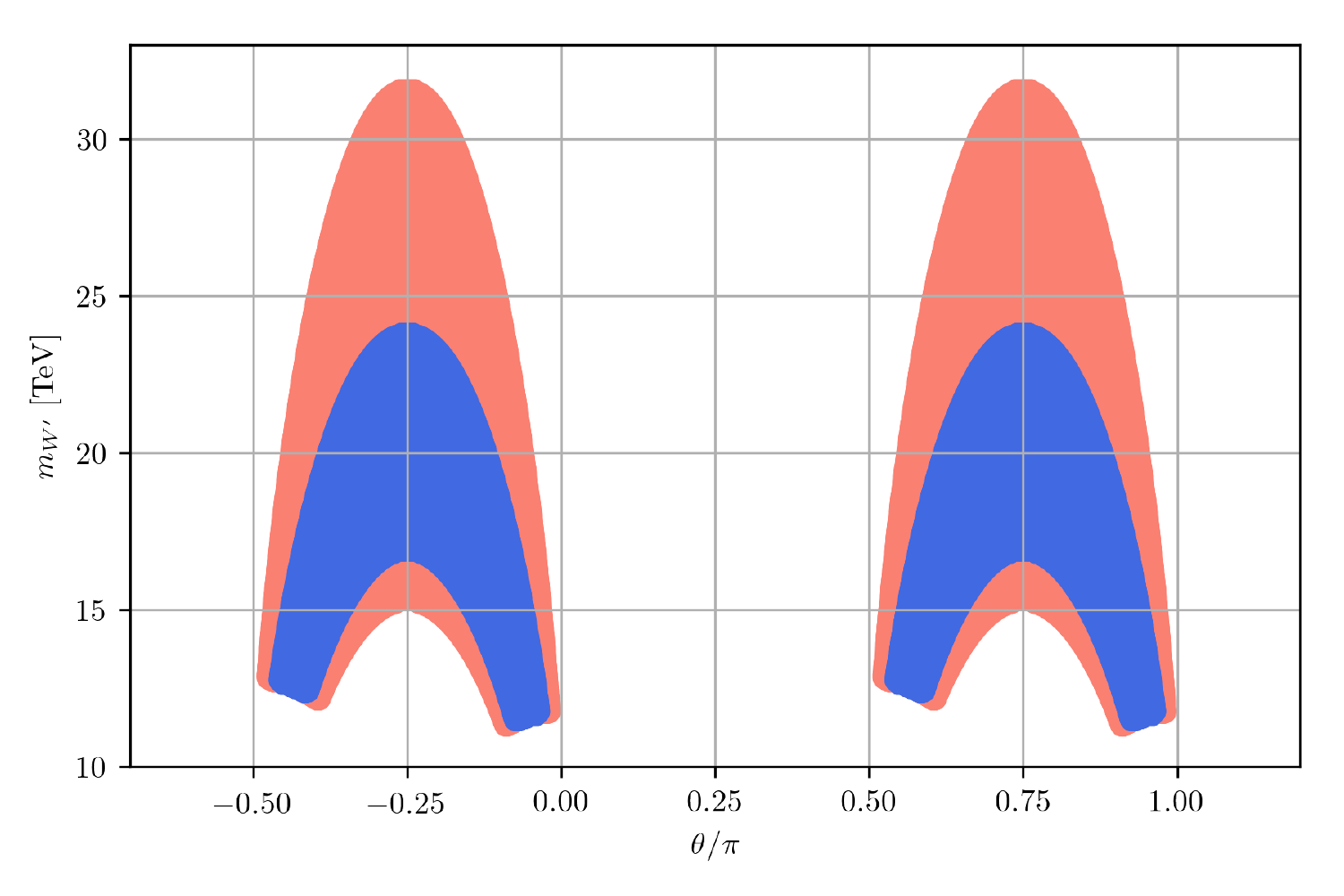}
    \caption{The favoured parameter regions compatible with the current experimental limits from 
${B^+}\to {K^+}\mu^- e^+$, ${B^+}\to {K^+} e^- \mu^+$. 
Shown are the 1$\sigma$ (blue) and 90\% confidence level (red) bands suggested by the measured $R_{K}$ and $R_{K*}$ ratios. 
}
    \label{fig:Allowable regions}
\end{figure}

A plot of the allowed model parameters is shown in Figure \ref{fig:Allowable regions}. From that figure it is clear that the favoured 
range of $\theta$ is approximately between $[-\frac{\pi}{2},0]$ or $[\frac{\pi}{2},\pi]$ and $m_{W'}/{\rm TeV}$ between $[12, 31]$. 
The identical nature of the two adjacent regions can be understood as follows.
Under the transformation
$\theta \to \theta+\pi$, $\sin\theta \to -\sin\theta, \cos\theta \to -\cos\theta$, and the leading order amplitudes for $b \to s\bar \ell \ell$ (which
are proportional to $\sin\theta \cos\theta$) are invariant. Also the amplitudes for the decay processes,  
${B^+}\to {K^+}\mu^- e^+$, ${B^+}\to {K^+} e^- \mu^+$, 
are proportional to $\sin^2\theta$ and $\cos^2\theta$ respectively, and 
are also invariant under $\theta \to \theta + \pi$.
It should be noted that the $R_{K^{(*)}}$ anomalies on their own can potentially have $m_{W'} < 12$ TeV, 
but the low mass cut-off is acquired due to the $B^{+} \to K^+ e^\mp \mu^\pm$ decay constraints. 

\begin{figure}[tpb!]
    \centering
    \begin{subfigure}{0.5\textwidth}
    \centering
        \includegraphics[width=1\linewidth]{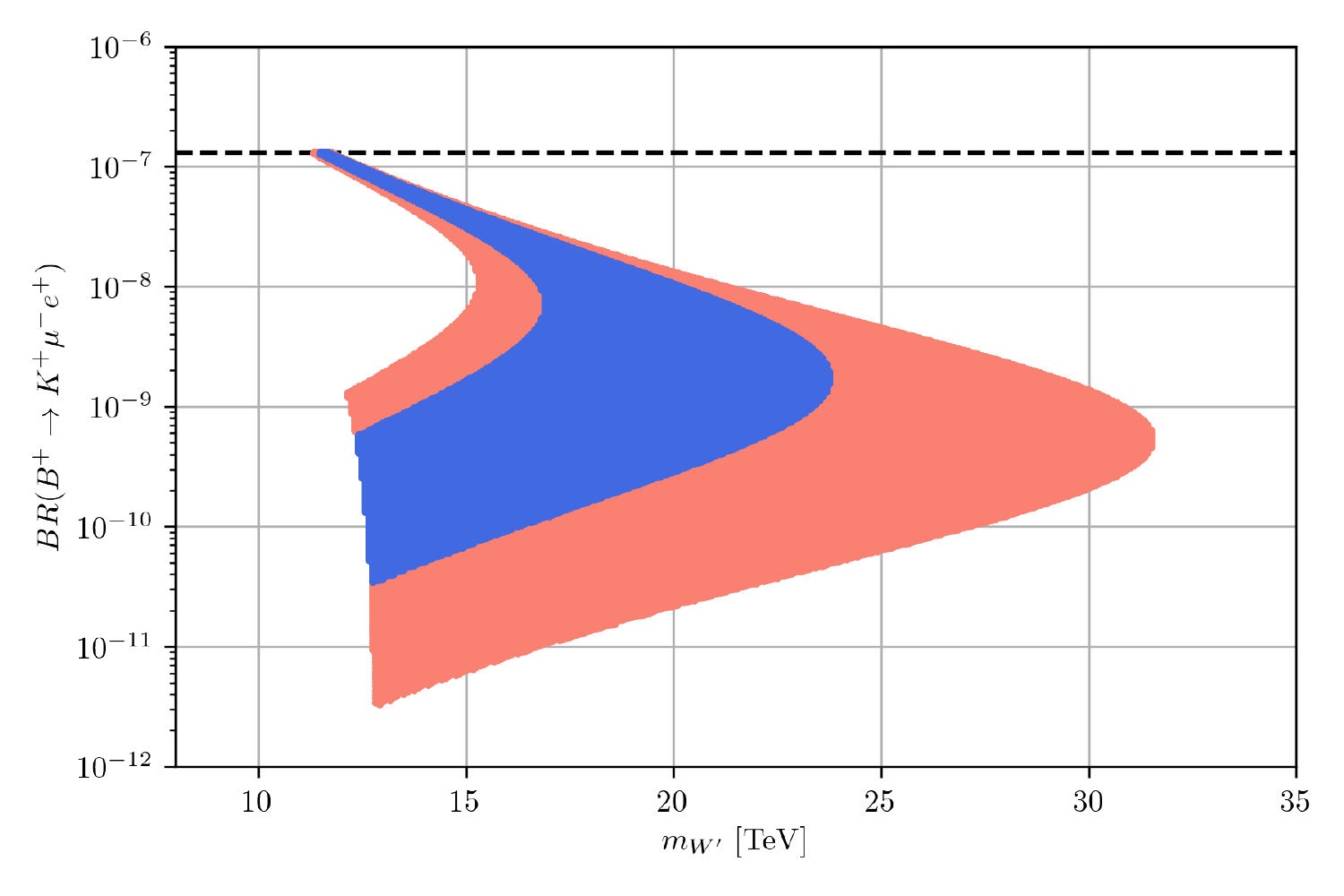}
        \caption{}
        \label{fig:BKmuePrediction}
    \end{subfigure}%
    \begin{subfigure}{0.5\textwidth}
    \centering
        \includegraphics[width=1.02\linewidth]{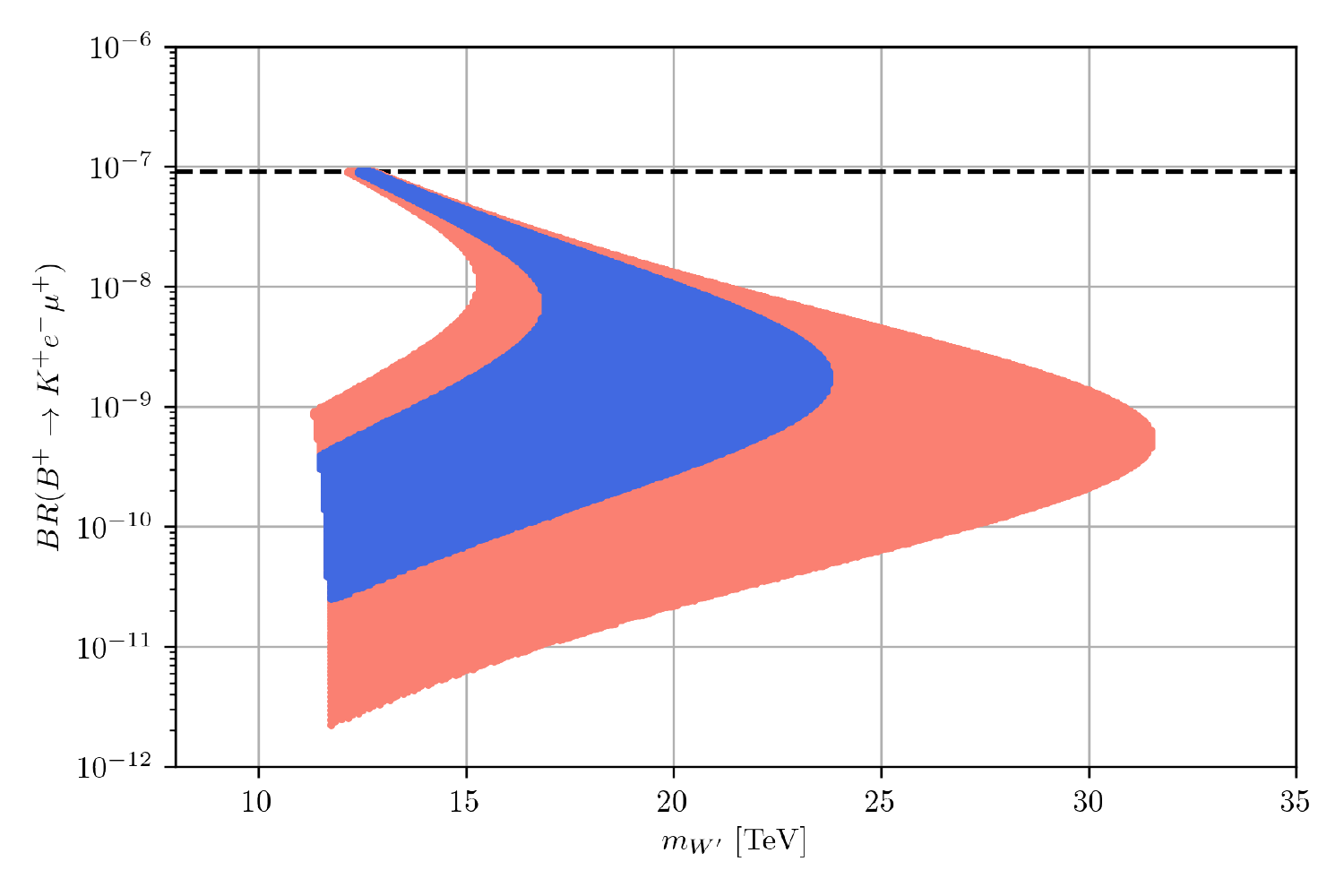}
        \caption{}
        \label{fig:BKemuPrediction}
    \end{subfigure}
    \caption{Expectation for (a) $BR({B^+}\to {K^+}\mu^- e^+)$ (b) $BR({B^+}\to {K^+} e^- \mu^+)$ 
 for the favoured parameter region identified in Figure \ref{fig:Allowable regions}. 
The black dashed lines correspond to the current experimental 90\% C.L. upper bounds on these branching fractions. 
}
    \label{fig:BR Predictions}
\end{figure}

For each point in the favoured region shown in Figure \ref{fig:Allowable regions}
we can calculate the
expected rates for the rare ${B^+}\to {K^+}\mu^- e^+$ and ${B^+}\to {K^+} e^- \mu^+$
processes. The result of this exercise is shown in 
Figure \ref{fig:BR Predictions}. 
Note that
$B^+ \to K^+\mu^- e^+$ probes $\sin^2\theta \approx 1$, while
$B^+ \to K^+\mu^+ e^-$ probes $\cos^2\theta \approx 1$, and thus these two decay channels are complimentary. Using the first $9\,\mathrm{fb}^{-1}$ LHCb is expected to be sensitive to the branching ratio of $B^+\to K^+e^\pm \mu^\mp$ at the level of $10^{-9}$ and scale almost linearly with integrated luminosity.\cite{Bediaga:2018lhg}

\begin{figure}[tbp!]
    \centering
    \begin{subfigure}{0.5\textwidth}
    \centering
        \includegraphics[width=1.01\linewidth]{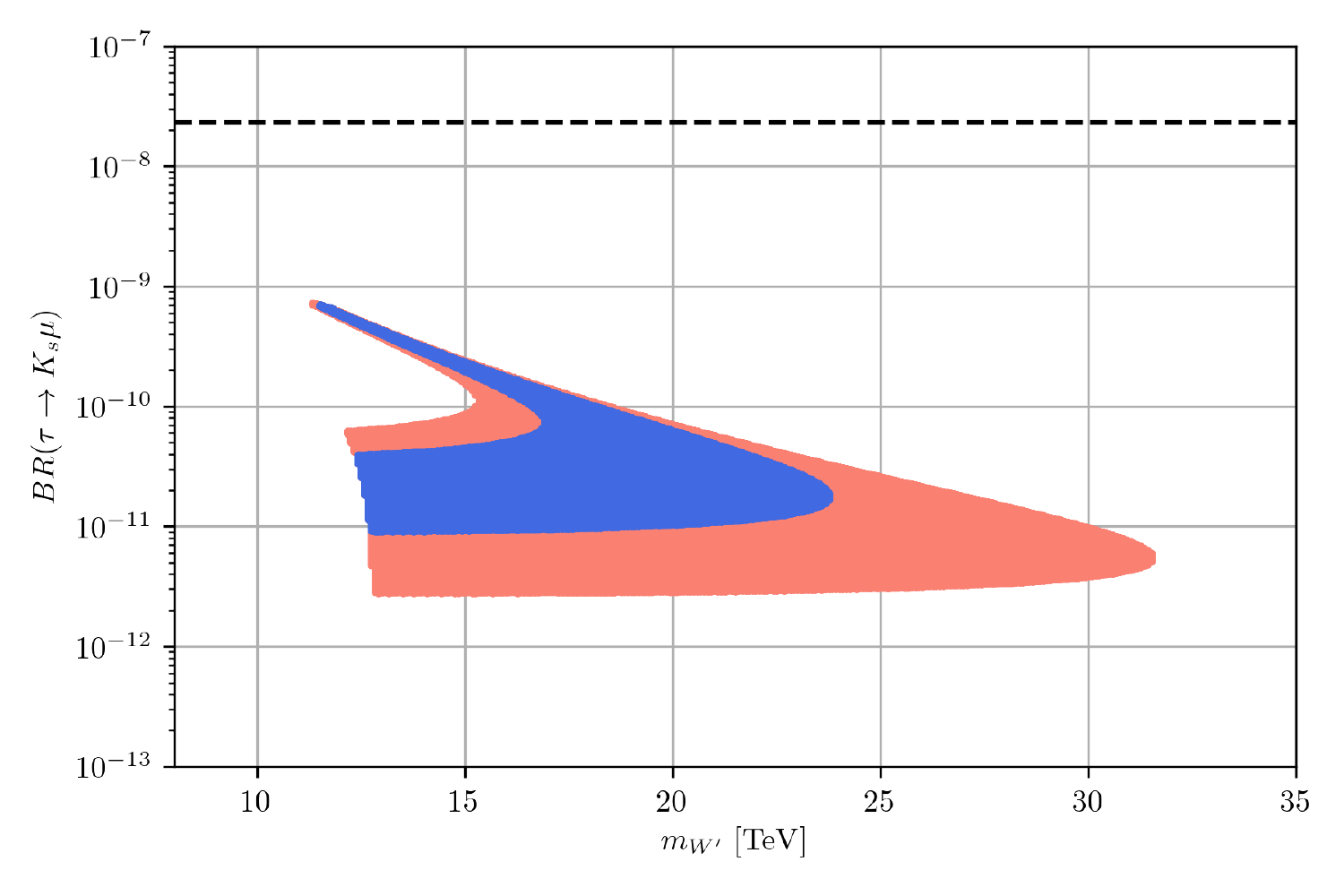}
        \caption{}
        \label{fig:tauKsmuPrediction2}
    \end{subfigure}%
    \begin{subfigure}{0.5\textwidth}
    \centering
        \includegraphics[width=1.01\linewidth]{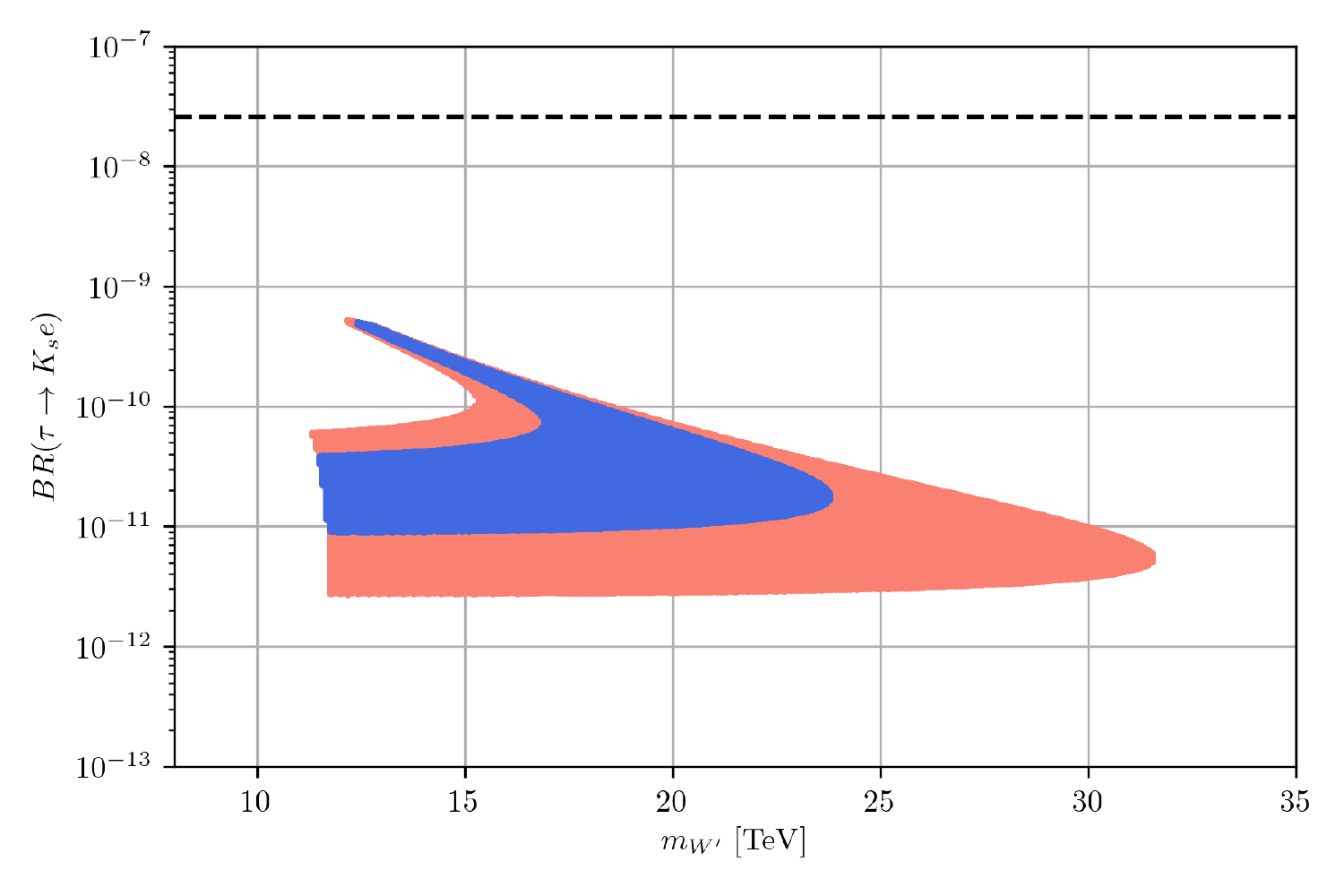}
        \caption{}
        \label{fig:tauKsePrediction2}
    \end{subfigure}
    \caption{Expectation for (a) $BR(\tau\to {K_s}\mu)$
 (b) $BR(\tau\to {K_s}e)$ for the favoured parameter region identified in Figure \ref{fig:Allowable regions}. 
The black dashed lines correspond to the current experimental 90\% C.L. upper bounds on these branching fractions. 
}
    \label{fig:BR Predictions2}
\end{figure}

In addition to further improvements to
$B^+ \to K^+ \mu^{\pm} e^{\mp}$
there
are a number of other ways to test this model. In the remainder of this paper we focus on making predictions for various rare decays that directly
involve the new physics invoked in explaining the $R_{K^{(*)}}$ anomalies.
We first consider the rare tau lepton decays: $\tau\to {K_s}\ell$, $\ell = e, \mu$.
The decay rate for the $\tau\to {K_s}\ell$ process is calculated to be
\begin{align}
\Gamma(\tau\to {K_s}\ell) = \frac{f_K^2 \alpha_s^2 \pi (m_\tau^2 - m_K^2)^2 [|K_{s\ell}|^2 |K_{d\tau}|^2 + |K_{s\tau}|^2|K_{d\ell}|^2]}{64 m_{W'}^4 m_{\tau}}
\ .
\end{align}
Here, $m_K \simeq 497.7$ MeV and $f_K \simeq 156.1$ MeV are the $K_s$ meson mass and decay constant respectively, and we have set the final state lepton mass to zero in the above
calculation.
With the ansatz, Eq.~(\ref{m5}), we have $K_{se} = \cos\theta$, $K_{s\mu} = \sin\theta$, $K_{d\tau} = 1$, $K_{d\ell}=0$.
Using the experimentally observed decay width, $\Gamma(\tau\to {\rm all}) \simeq 2.27\times10^{-12}$ GeV,  
the branching fraction, 
$BR(\tau\to {K_s}\ell) = \Gamma(\tau\to K_s\ell)/\Gamma(\tau\to {\rm all})$, can
then be obtained.
Our results are shown in 
Figure \ref{fig:BR Predictions2}. The Belle II experiment will search for $\tau\to K_s\ell$ decays with an improved sensitivity of $5\times 10^{-10}$ ($4\times 10^{-10}$) for $\tau\to K_s e$ ($\tau\to K_s\mu$).\cite{Kou:2018nap}

\begin{figure}[tbp!]
    \centering
    \begin{subfigure}{0.5\textwidth}
    \centering
        \includegraphics[width=1\linewidth]{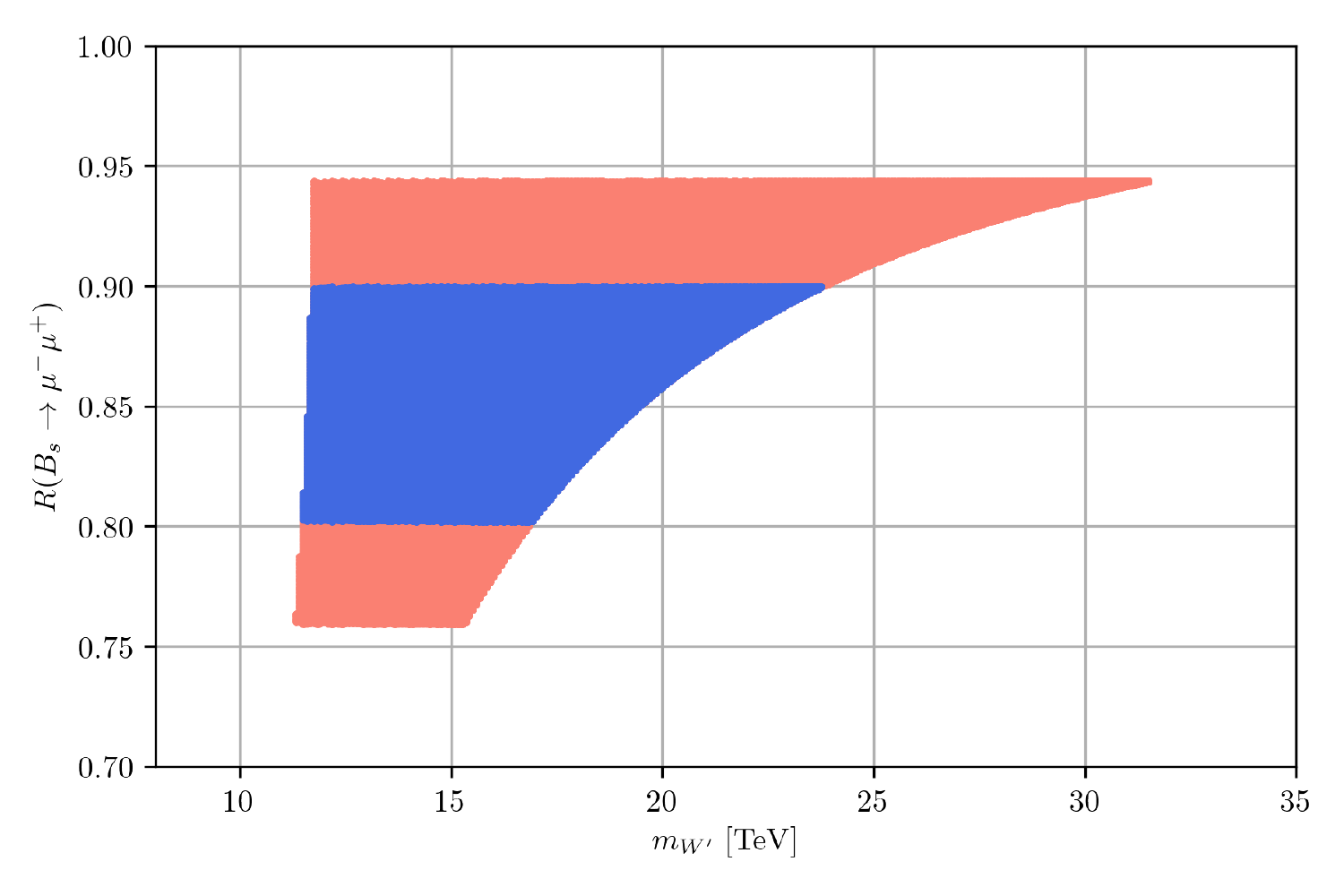}
        \caption{}
        \label{fig:BsmumuPrediction}
    \end{subfigure}%
    \begin{subfigure}{0.5\textwidth}
    \centering
        \includegraphics[width=1.02\linewidth]{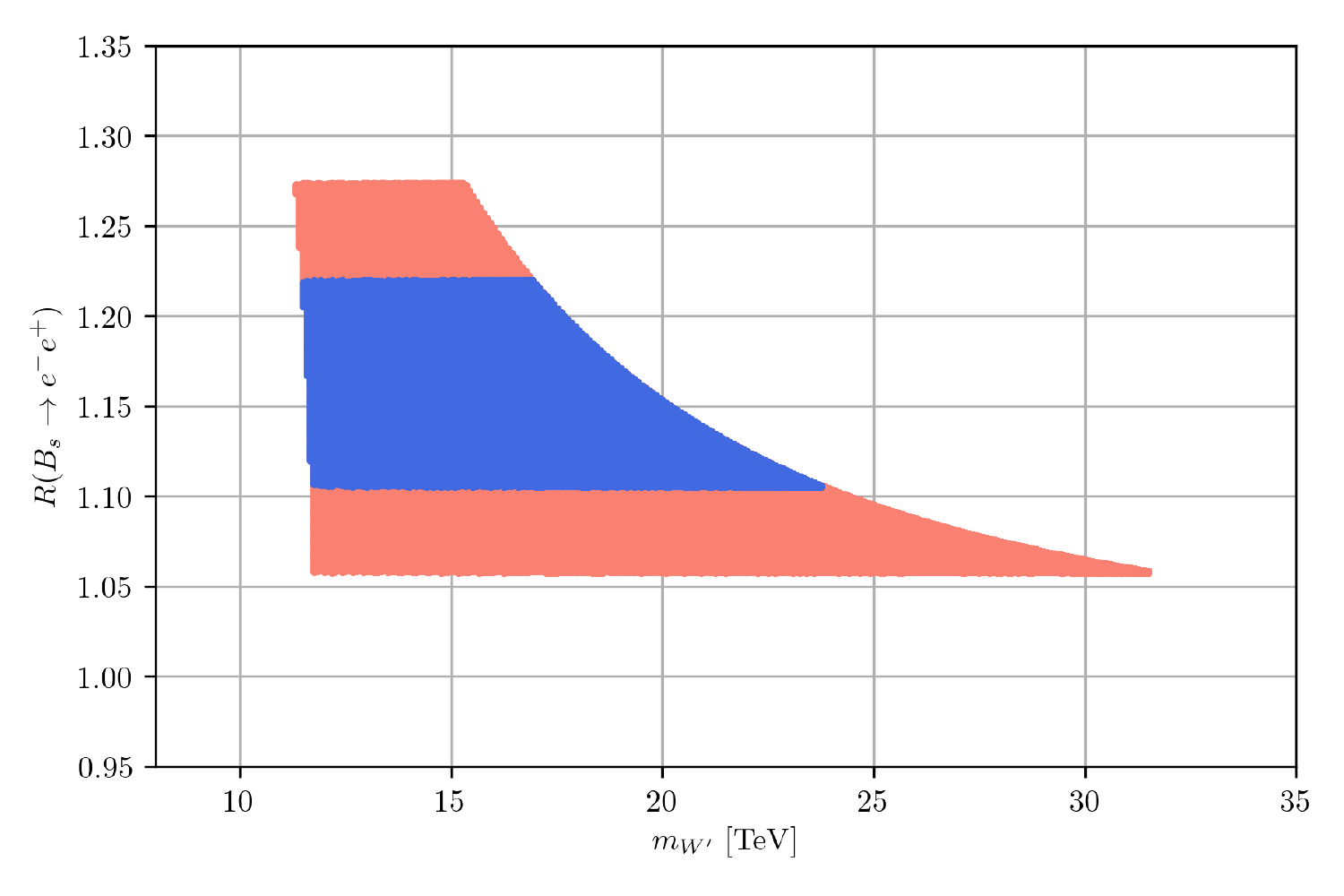}
        \caption{}
        \label{fig:BseePrediction}
    \end{subfigure}
    \begin{subfigure}{0.5\textwidth}
    \centering
        \includegraphics[width=1.01\linewidth]{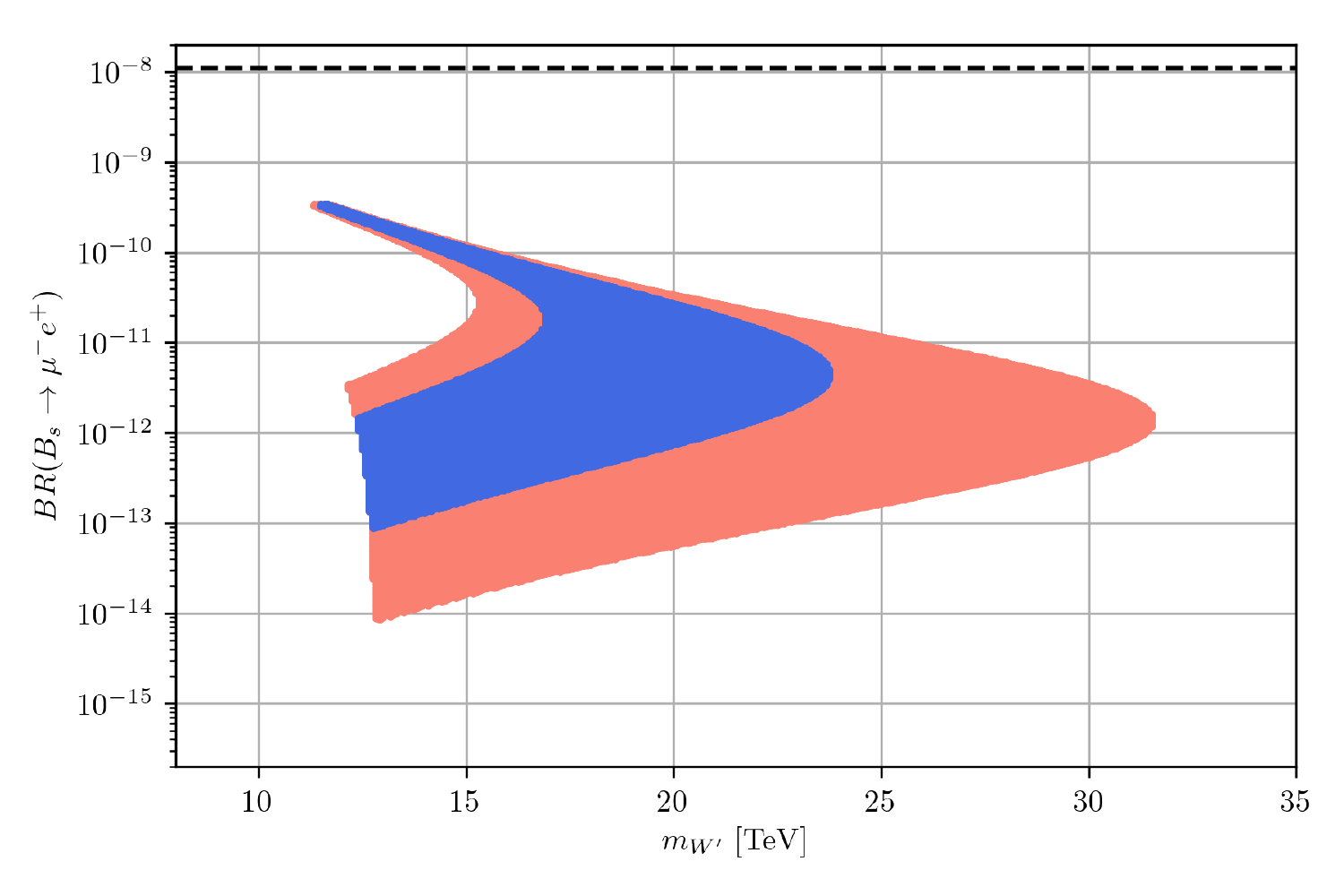}
        \caption{}
        \label{fig:BsmuePrediction}
    \end{subfigure}%
    \begin{subfigure}{0.5\textwidth}
    \centering
        \includegraphics[width=1.01\linewidth]{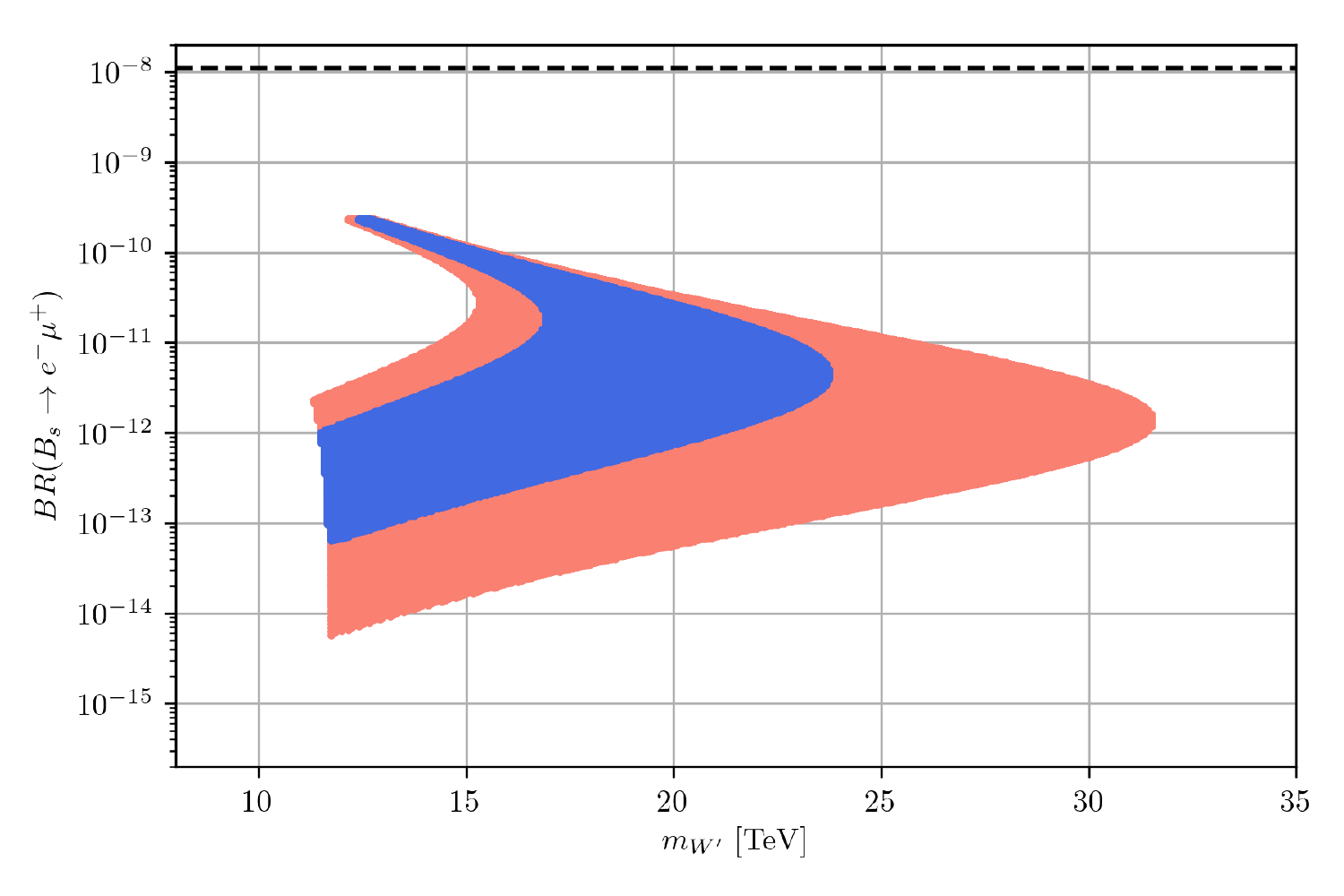}
        \caption{}
        \label{fig:BsemuPrediction}
    \end{subfigure}
    \caption{Expectation for (a) $R(B_s\to\mu^-\mu^+)$  (b) $R(B_s\to e^- e^+)$ (c) $BR(B_s\to\mu^- e^+)$ (d) $BR(B_s\to e^-\mu^+)$ for the favoured region
of parameter space identified in Figure  \ref{fig:Allowable regions}. 
}
    \label{fig:Bs Predictions}
\end{figure}

The effective Lagrangian that induces modifications to the $R_K$ ratio also modifies
the two-body $B_s$ decays:
$B_s\to\mu^-\mu^+$ and $B_s\to e^-e^+$.
These decays also arise in the standard model, and so it is useful to compute the ratio
\begin{equation}
	 R(B_s\to\ell^- \ell^+)
\equiv \frac{\Gamma (B_s\to\ell^-\ell^+)}{\Gamma_{SM}(B_s\to\ell^-\ell^+)}
\end{equation}
where the numerator, $\Gamma (B_s \to \ell^- \ell^+)$, includes the new physics ($W'$) contributions as well as the standard model contribution.
In this model we expect $R(B_s\to\mu^- \mu^+) \simeq (1+R_K)/2$, 
and $R(B_s\to e^- e^+) \simeq (3-R_K)/2$. 
In 
Figure \ref{fig:Bs Predictions} 
we have calculated the predictions for  $R(B_s\to\ell^- \ell^+)$. 
A comparison of the experimental values~\cite{Tanabashi:2018oca} with the SM predictions~\cite{Bobeth:2013uxa} shows that the $R(B_s \to \mu^- \mu^+)$ 
ratio inferred from measurement is $R(B_s\to \mu^-\mu^+) = 0.7\pm0.3$. This value is
consistent with what we would expect given the central values of $R_K$ and $R_{K^*}$, but of course the current error is too large to rigorously test this model. 
In Figure \ref{fig:Bs Predictions} 
we have also
shown the predicted branching ratios $BR(B_s \to \mu^- e^+)$ and $BR(B_s \to e^- \mu^+)$, together with the 90\% C.L.~upper bound $BR(B_s \to e^\pm \mu^\mp) < 1.1\times 10^{-8}$.

	The vector leptoquark also modifies the two lepton univerality ratios $R_D^{\mu/e} = \Gamma(B\to D \mu \bar\nu)/\Gamma(B\to D e\bar\nu)$ and $R_{D^*}^{e/\mu}=\Gamma(B\to D^* e\bar\nu)/\Gamma(B\to D^*\mu\bar\nu)$ via its couplings to up-type quarks and neutrinos. These ratios have been measured by the Belle experiment: $R_D^{\mu/e}=0.995\pm0.022\pm0.039$~\cite{Glattauer:2015teq} and $R_{D^*}^{e/\mu}=1.04\pm0.05\pm0.01$~\cite{Abdesselam:2017kjf}, where the first and second uncertainties are statistical and systematic respectively. To leading order in the contribution of the vector leptoquark the lepton universality ratios are given by
	\begin{align}
		R_{D}^{\mu/e} &
		\simeq 
		R_{D,SM}^{\mu/e}\left(1+ \frac{\sqrt{2}\pi \alpha_s \cos\theta_c \sin2\theta }{ V_{cb} G_F m_{W'}^2} \right)\;, 
	        &
		R_{D^*}^{e/\mu} &
		\simeq 
		R_{D^*,SM}^{e/\mu} \left(1- \frac{\sqrt{2}\pi\alpha_s \cos\theta_c  \sin2\theta }{V_{cb} G_F m_{W'}^2} \right)\;,
	\end{align}
	where $\theta_c$ denotes the Cabibbo angle. For the region of interest the deviation from the SM value is about one order of magnitude smaller than the experimental sensitivity of Belle and hence does not currently pose a new constraint. 

We have briefly looked at the $\mu \to e \gamma$ radiative decay. This decay arises at one-loop level, with virtual down-type quarks and $W'$ gauge boson propagators in the loop.
Making use of the general calculation given in Ref.~\cite{Lavoura:2003xp}, we show that the first two terms in the $m_b^2/m_{W'}^2$ expansion vanish: the first one due to unitarity and the second one 
\begin{align}
\Gamma(\mu\to e\gamma) \simeq \frac{9\,\alpha_{em} \alpha_s^2 m_b^4 m_{\mu}^5 \left(2Q_b + Q_{W'}\right)^2 \sin^2\theta \cos^2\theta}{256 m_{W'}^8}
\end{align}
is proportional to $(2Q_b+Q_{W'})^2$ and thus vanishes as the charge assignments in this model satisfy $Q_b = -1/3$ and $Q_{W'} = 2/3$. 
Hence we do not expect the $\mu \to e \gamma$ process to be important in this model. 

A similar conclusion holds for $\mu\to eee$ and $\mu\to e$ conversion in nuclei, because due to dipole dominance the decay width $\Gamma(\mu\to eee)$ and the conversion rate $CR(\mu N\to eN)$ are directly proportional to $\Gamma(\mu\to e\gamma)$. In particular, there are no tree-level contributions to $\mu\to e$ conversion for the $K$ matrix in Eq.~\eqref{m5}.

\section{Conclusion}
\label{sec:conclusions}
We have proposed a Pati-Salam variant $SU(4)$ theory, with gauge group $SU(4)_C\times SU(2)_L\times U(1)_{Y'}$, which is capable of explaining the $R_K$ and $R_{K^*}$ anomalies 
via new gauge interactions.
The model is consistent with experimental constraints, including
the stringent limits on ${B^+}\to {K^+}\mu^- e^+$ and ${B^+}\to {K^+} e^- \mu^+$ decays. 
In this model, the chiral left-handed fermions are arranged in a similar fashion to the original Pati-Salam model, i.e.~with leptons making up the
fourth colour, while the chiral right-handed fermions are treated quite differently. 
The model 
features $SU(4)$ symmetry breaking via the introduction of a $SU(4)$ scalar multiplet $\chi$ with a VEV $w \gtrsim 10$ TeV and electroweak symmetry breaking 
via scalars $\phi$ and $\Delta$ with VEVs that satisfy $\sqrt{v^2 + u^2} \simeq 174 \ {\rm GeV}$. In addition to new scalar particles, the model contains  
new charged ($\frac{2}{3} e$) $W'$ and neutral $Z'$ gauge bosons along with heavy exotic charged $E_{L,R}^{-}$ and neutral $N_{L,R}$ fermions.
The charged leptoquark gauge bosons $W'$ couple in a chiral manner to the familiar quarks and leptons and can thereby 
interfere with SM weak processes. The theory makes predictions for ${B^+}\to {K^+}\mu^- e^+$, ${B^+}\to {K^+} e^- \mu^+$, $\tau\to {K_s}\ell$, 
$B_s\to\mu^-\mu^+$, as well as the highly suppressed  $B_s\to\mu^- e^+$ and  $B_s\to e^- \mu^+$ processes. For instance, for the leptonic $B_s \to \mu^- \mu^+$ decay channel the rate is predicted to satisfy: $\Gamma (B_s \to \mu^- \mu^+)/\Gamma_{SM} (B_s \to \mu^- \mu^+) = (1 + R_K)/2$.
These predictions 
can be tested  at the LHCb and Belle II experiments when increased statistics become available.

The leptoquark gauge boson phenomenology of the chiral $SU(4)$ Pati-Salam model considered
will be relevant for more general chiral $SU(4)$ models. In particular, the model can easily be 
extended to the full Pati-Salam gauge group: $SU(4)\otimes SU(2)_L \otimes SU(2)_R$. In this 
case, the three $SU(4)$ singlet fermions in Table~\ref{tab:particles} unify into a $SU(2)_R$ triplet, that is the 
fermion content of each generation have gauge transformation: 
$Q_L \sim (4,2,1), Q_R \sim (4,1,2), F_R \sim (1,1,3)$. 
The $SU(4)$ leptoquark gauge bosons of such extended models can explain the measured $R_K$ 
deviations in the same manner as discussed here. However, since such models typically require more 
scalar degrees of freedom, there are more observable signatures of new physics, including the 
possibility of explaining the $R_D$ anomalies via scalar leptoquarks. Although very interesting and 
topical in light of the tantalizing experimental hints, we leave further investigations along these lines 
for future work.

\section*{Acknowledgements}
This work has been supported in part by the Australian Research Council.

\setlength{\bibsep}{0pt}
\bibliography{refs}

\begin{thebibliography}{49}%
\makeatletter
\providecommand \@ifxundefined [1]{%
 \@ifx{#1\undefined}
}%
\providecommand \@ifnum [1]{%
 \ifnum #1\expandafter \@firstoftwo
 \else \expandafter \@secondoftwo
 \fi
}%
\providecommand \@ifx [1]{%
 \ifx #1\expandafter \@firstoftwo
 \else \expandafter \@secondoftwo
 \fi
}%
\providecommand \natexlab [1]{#1}%
\providecommand \enquote  [1]{``#1''}%
\providecommand \bibnamefont  [1]{#1}%
\providecommand \bibfnamefont [1]{#1}%
\providecommand \citenamefont [1]{#1}%
\providecommand \href@noop [0]{\@secondoftwo}%
\providecommand \href [0]{\begingroup \@sanitize@url \@href}%
\providecommand \@href[1]{\@@startlink{#1}\@@href}%
\providecommand \@@href[1]{\endgroup#1\@@endlink}%
\providecommand \@sanitize@url [0]{\catcode `\\12\catcode `\$12\catcode
  `\&12\catcode `\#12\catcode `\^12\catcode `\_12\catcode `\%12\relax}%
\providecommand \@@startlink[1]{}%
\providecommand \@@endlink[0]{}%
\providecommand \url  [0]{\begingroup\@sanitize@url \@url }%
\providecommand \@url [1]{\endgroup\@href {#1}{\urlprefix }}%
\providecommand \urlprefix  [0]{URL }%
\providecommand \Eprint [0]{\href }%
\providecommand \doibase [0]{http://dx.doi.org/}%
\providecommand \selectlanguage [0]{\@gobble}%
\providecommand \bibinfo  [0]{\@secondoftwo}%
\providecommand \bibfield  [0]{\@secondoftwo}%
\providecommand \translation [1]{[#1]}%
\providecommand \BibitemOpen [0]{}%
\providecommand \bibitemStop [0]{}%
\providecommand \bibitemNoStop [0]{.\EOS\space}%
\providecommand \EOS [0]{\spacefactor3000\relax}%
\providecommand \BibitemShut  [1]{\csname bibitem#1\endcsname}%
\let\auto@bib@innerbib\@empty
\bibitem [{\citenamefont {Aaij}\ \emph
  {et~al.}(2014{\natexlab{a}})\citenamefont {Aaij} \emph
  {et~al.}}]{Aaij:2014ora}%
  \BibitemOpen
  \bibfield  {author} {\bibinfo {author} {\bibfnamefont {R.}~\bibnamefont
  {Aaij}} \emph {et~al.} (\bibinfo {collaboration} {LHCb}),\ }\href {\doibase
  10.1103/PhysRevLett.113.151601} {\bibfield  {journal} {\bibinfo  {journal}
  {Phys. Rev. Lett.}\ }\textbf {\bibinfo {volume} {113}},\ \bibinfo {pages}
  {151601} (\bibinfo {year} {2014}{\natexlab{a}})},\ \Eprint
  {http://arxiv.org/abs/1406.6482} {arXiv:1406.6482 [hep-ex]} \BibitemShut
  {NoStop}%
\bibitem [{\citenamefont {Aaij}\ \emph {et~al.}(2017)\citenamefont {Aaij} \emph
  {et~al.}}]{Aaij:2017vbb}%
  \BibitemOpen
  \bibfield  {author} {\bibinfo {author} {\bibfnamefont {R.}~\bibnamefont
  {Aaij}} \emph {et~al.} (\bibinfo {collaboration} {LHCb}),\ }\href {\doibase
  10.1007/JHEP08(2017)055} {\bibfield  {journal} {\bibinfo  {journal} {JHEP}\
  }\textbf {\bibinfo {volume} {08}},\ \bibinfo {pages} {055} (\bibinfo {year}
  {2017})},\ \Eprint {http://arxiv.org/abs/1705.05802} {arXiv:1705.05802
  [hep-ex]} \BibitemShut {NoStop}%
\bibitem [{\citenamefont {Aaij}\ \emph {et~al.}(2013)\citenamefont {Aaij} \emph
  {et~al.}}]{Aaij:2013qta}%
  \BibitemOpen
  \bibfield  {author} {\bibinfo {author} {\bibfnamefont {R.}~\bibnamefont
  {Aaij}} \emph {et~al.} (\bibinfo {collaboration} {LHCb}),\ }\href {\doibase
  10.1103/PhysRevLett.111.191801} {\bibfield  {journal} {\bibinfo  {journal}
  {Phys. Rev. Lett.}\ }\textbf {\bibinfo {volume} {111}},\ \bibinfo {pages}
  {191801} (\bibinfo {year} {2013})},\ \Eprint {http://arxiv.org/abs/1308.1707}
  {arXiv:1308.1707 [hep-ex]} \BibitemShut {NoStop}%
\bibitem [{\citenamefont {Aaij}\ \emph {et~al.}(2016)\citenamefont {Aaij} \emph
  {et~al.}}]{Aaij:2015oid}%
  \BibitemOpen
  \bibfield  {author} {\bibinfo {author} {\bibfnamefont {R.}~\bibnamefont
  {Aaij}} \emph {et~al.} (\bibinfo {collaboration} {LHCb}),\ }\href {\doibase
  10.1007/JHEP02(2016)104} {\bibfield  {journal} {\bibinfo  {journal} {JHEP}\
  }\textbf {\bibinfo {volume} {02}},\ \bibinfo {pages} {104} (\bibinfo {year}
  {2016})},\ \Eprint {http://arxiv.org/abs/1512.04442} {arXiv:1512.04442
  [hep-ex]} \BibitemShut {NoStop}%
\bibitem [{\citenamefont {Wehle}\ \emph {et~al.}(2017)\citenamefont {Wehle}
  \emph {et~al.}}]{Wehle:2016yoi}%
  \BibitemOpen
  \bibfield  {author} {\bibinfo {author} {\bibfnamefont {S.}~\bibnamefont
  {Wehle}} \emph {et~al.} (\bibinfo {collaboration} {Belle}),\ }\href {\doibase
  10.1103/PhysRevLett.118.111801} {\bibfield  {journal} {\bibinfo  {journal}
  {Phys. Rev. Lett.}\ }\textbf {\bibinfo {volume} {118}},\ \bibinfo {pages}
  {111801} (\bibinfo {year} {2017})},\ \Eprint
  {http://arxiv.org/abs/1612.05014} {arXiv:1612.05014 [hep-ex]} \BibitemShut
  {NoStop}%
\bibitem [{\citenamefont {Aaij}\ \emph
  {et~al.}(2014{\natexlab{b}})\citenamefont {Aaij} \emph
  {et~al.}}]{Aaij:2014pli}%
  \BibitemOpen
  \bibfield  {author} {\bibinfo {author} {\bibfnamefont {R.}~\bibnamefont
  {Aaij}} \emph {et~al.} (\bibinfo {collaboration} {LHCb}),\ }\href {\doibase
  10.1007/JHEP06(2014)133} {\bibfield  {journal} {\bibinfo  {journal} {JHEP}\
  }\textbf {\bibinfo {volume} {06}},\ \bibinfo {pages} {133} (\bibinfo {year}
  {2014}{\natexlab{b}})},\ \Eprint {http://arxiv.org/abs/1403.8044}
  {arXiv:1403.8044 [hep-ex]} \BibitemShut {NoStop}%
\bibitem [{\citenamefont {Aaij}\ \emph {et~al.}(2015)\citenamefont {Aaij} \emph
  {et~al.}}]{Aaij:2015esa}%
  \BibitemOpen
  \bibfield  {author} {\bibinfo {author} {\bibfnamefont {R.}~\bibnamefont
  {Aaij}} \emph {et~al.} (\bibinfo {collaboration} {LHCb}),\ }\href {\doibase
  10.1007/JHEP09(2015)179} {\bibfield  {journal} {\bibinfo  {journal} {JHEP}\
  }\textbf {\bibinfo {volume} {09}},\ \bibinfo {pages} {179} (\bibinfo {year}
  {2015})},\ \Eprint {http://arxiv.org/abs/1506.08777} {arXiv:1506.08777
  [hep-ex]} \BibitemShut {NoStop}%
\bibitem [{\citenamefont {Hiller}\ and\ \citenamefont
  {Kruger}(2004)}]{Hiller:2003js}%
  \BibitemOpen
  \bibfield  {author} {\bibinfo {author} {\bibfnamefont {G.}~\bibnamefont
  {Hiller}}\ and\ \bibinfo {author} {\bibfnamefont {F.}~\bibnamefont
  {Kruger}},\ }\href {\doibase 10.1103/PhysRevD.69.074020} {\bibfield
  {journal} {\bibinfo  {journal} {Phys. Rev.}\ }\textbf {\bibinfo {volume}
  {D69}},\ \bibinfo {pages} {074020} (\bibinfo {year} {2004})},\ \Eprint
  {http://arxiv.org/abs/hep-ph/0310219} {arXiv:hep-ph/0310219 [hep-ph]}
  \BibitemShut {NoStop}%
\bibitem [{\citenamefont {Capdevila}\ \emph {et~al.}(2016)\citenamefont
  {Capdevila}, \citenamefont {Descotes-Genon}, \citenamefont {Matias},\ and\
  \citenamefont {Virto}}]{Capdevila:2016ivx}%
  \BibitemOpen
  \bibfield  {author} {\bibinfo {author} {\bibfnamefont {B.}~\bibnamefont
  {Capdevila}}, \bibinfo {author} {\bibfnamefont {S.}~\bibnamefont
  {Descotes-Genon}}, \bibinfo {author} {\bibfnamefont {J.}~\bibnamefont
  {Matias}}, \ and\ \bibinfo {author} {\bibfnamefont {J.}~\bibnamefont
  {Virto}},\ }\href {\doibase 10.1007/JHEP10(2016)075} {\bibfield  {journal}
  {\bibinfo  {journal} {JHEP}\ }\textbf {\bibinfo {volume} {10}},\ \bibinfo
  {pages} {075} (\bibinfo {year} {2016})},\ \Eprint
  {http://arxiv.org/abs/1605.03156} {arXiv:1605.03156 [hep-ph]} \BibitemShut
  {NoStop}%
\bibitem [{\citenamefont {Capdevila}\ \emph {et~al.}(2018)\citenamefont
  {Capdevila}, \citenamefont {Crivellin}, \citenamefont {Descotes-Genon},
  \citenamefont {Matias},\ and\ \citenamefont {Virto}}]{Capdevila:2017bsm}%
  \BibitemOpen
  \bibfield  {author} {\bibinfo {author} {\bibfnamefont {B.}~\bibnamefont
  {Capdevila}}, \bibinfo {author} {\bibfnamefont {A.}~\bibnamefont
  {Crivellin}}, \bibinfo {author} {\bibfnamefont {S.}~\bibnamefont
  {Descotes-Genon}}, \bibinfo {author} {\bibfnamefont {J.}~\bibnamefont
  {Matias}}, \ and\ \bibinfo {author} {\bibfnamefont {J.}~\bibnamefont
  {Virto}},\ }\href {\doibase 10.1007/JHEP01(2018)093} {\bibfield  {journal}
  {\bibinfo  {journal} {JHEP}\ }\textbf {\bibinfo {volume} {01}},\ \bibinfo
  {pages} {093} (\bibinfo {year} {2018})},\ \Eprint
  {http://arxiv.org/abs/1704.05340} {arXiv:1704.05340 [hep-ph]} \BibitemShut
  {NoStop}%
\bibitem [{\citenamefont {Altmannshofer}\ \emph {et~al.}(2017)\citenamefont
  {Altmannshofer}, \citenamefont {Stangl},\ and\ \citenamefont
  {Straub}}]{Altmannshofer:2017yso}%
  \BibitemOpen
  \bibfield  {author} {\bibinfo {author} {\bibfnamefont {W.}~\bibnamefont
  {Altmannshofer}}, \bibinfo {author} {\bibfnamefont {P.}~\bibnamefont
  {Stangl}}, \ and\ \bibinfo {author} {\bibfnamefont {D.~M.}\ \bibnamefont
  {Straub}},\ }\href {\doibase 10.1103/PhysRevD.96.055008} {\bibfield
  {journal} {\bibinfo  {journal} {Phys. Rev.}\ }\textbf {\bibinfo {volume}
  {D96}},\ \bibinfo {pages} {055008} (\bibinfo {year} {2017})},\ \Eprint
  {http://arxiv.org/abs/1704.05435} {arXiv:1704.05435 [hep-ph]} \BibitemShut
  {NoStop}%
\bibitem [{\citenamefont {D'Amico}\ \emph {et~al.}(2017)\citenamefont
  {D'Amico}, \citenamefont {Nardecchia}, \citenamefont {Panci}, \citenamefont
  {Sannino}, \citenamefont {Strumia}, \citenamefont {Torre},\ and\
  \citenamefont {Urbano}}]{DAmico:2017mtc}%
  \BibitemOpen
  \bibfield  {author} {\bibinfo {author} {\bibfnamefont {G.}~\bibnamefont
  {D'Amico}}, \bibinfo {author} {\bibfnamefont {M.}~\bibnamefont {Nardecchia}},
  \bibinfo {author} {\bibfnamefont {P.}~\bibnamefont {Panci}}, \bibinfo
  {author} {\bibfnamefont {F.}~\bibnamefont {Sannino}}, \bibinfo {author}
  {\bibfnamefont {A.}~\bibnamefont {Strumia}}, \bibinfo {author} {\bibfnamefont
  {R.}~\bibnamefont {Torre}}, \ and\ \bibinfo {author} {\bibfnamefont
  {A.}~\bibnamefont {Urbano}},\ }\href {\doibase 10.1007/JHEP09(2017)010}
  {\bibfield  {journal} {\bibinfo  {journal} {JHEP}\ }\textbf {\bibinfo
  {volume} {09}},\ \bibinfo {pages} {010} (\bibinfo {year} {2017})},\ \Eprint
  {http://arxiv.org/abs/1704.05438} {arXiv:1704.05438 [hep-ph]} \BibitemShut
  {NoStop}%
\bibitem [{\citenamefont {Hiller}\ and\ \citenamefont
  {Nisandzic}(2017)}]{Hiller:2017bzc}%
  \BibitemOpen
  \bibfield  {author} {\bibinfo {author} {\bibfnamefont {G.}~\bibnamefont
  {Hiller}}\ and\ \bibinfo {author} {\bibfnamefont {I.}~\bibnamefont
  {Nisandzic}},\ }\href {\doibase 10.1103/PhysRevD.96.035003} {\bibfield
  {journal} {\bibinfo  {journal} {Phys. Rev.}\ }\textbf {\bibinfo {volume}
  {D96}},\ \bibinfo {pages} {035003} (\bibinfo {year} {2017})},\ \Eprint
  {http://arxiv.org/abs/1704.05444} {arXiv:1704.05444 [hep-ph]} \BibitemShut
  {NoStop}%
\bibitem [{\citenamefont {Geng}\ \emph {et~al.}(2017)\citenamefont {Geng},
  \citenamefont {Grinstein}, \citenamefont {Jäger}, \citenamefont
  {Martin~Camalich}, \citenamefont {Ren},\ and\ \citenamefont
  {Shi}}]{Geng:2017svp}%
  \BibitemOpen
  \bibfield  {author} {\bibinfo {author} {\bibfnamefont {L.-S.}\ \bibnamefont
  {Geng}}, \bibinfo {author} {\bibfnamefont {B.}~\bibnamefont {Grinstein}},
  \bibinfo {author} {\bibfnamefont {S.}~\bibnamefont {Jäger}}, \bibinfo
  {author} {\bibfnamefont {J.}~\bibnamefont {Martin~Camalich}}, \bibinfo
  {author} {\bibfnamefont {X.-L.}\ \bibnamefont {Ren}}, \ and\ \bibinfo
  {author} {\bibfnamefont {R.-X.}\ \bibnamefont {Shi}},\ }\href {\doibase
  10.1103/PhysRevD.96.093006} {\bibfield  {journal} {\bibinfo  {journal} {Phys.
  Rev.}\ }\textbf {\bibinfo {volume} {D96}},\ \bibinfo {pages} {093006}
  (\bibinfo {year} {2017})},\ \Eprint {http://arxiv.org/abs/1704.05446}
  {arXiv:1704.05446 [hep-ph]} \BibitemShut {NoStop}%
\bibitem [{\citenamefont {Ciuchini}\ \emph {et~al.}(2017)\citenamefont
  {Ciuchini}, \citenamefont {Coutinho}, \citenamefont {Fedele}, \citenamefont
  {Franco}, \citenamefont {Paul}, \citenamefont {Silvestrini},\ and\
  \citenamefont {Valli}}]{Ciuchini:2017mik}%
  \BibitemOpen
  \bibfield  {author} {\bibinfo {author} {\bibfnamefont {M.}~\bibnamefont
  {Ciuchini}}, \bibinfo {author} {\bibfnamefont {A.~M.}\ \bibnamefont
  {Coutinho}}, \bibinfo {author} {\bibfnamefont {M.}~\bibnamefont {Fedele}},
  \bibinfo {author} {\bibfnamefont {E.}~\bibnamefont {Franco}}, \bibinfo
  {author} {\bibfnamefont {A.}~\bibnamefont {Paul}}, \bibinfo {author}
  {\bibfnamefont {L.}~\bibnamefont {Silvestrini}}, \ and\ \bibinfo {author}
  {\bibfnamefont {M.}~\bibnamefont {Valli}},\ }\href {\doibase
  10.1140/epjc/s10052-017-5270-2} {\bibfield  {journal} {\bibinfo  {journal}
  {Eur. Phys. J.}\ }\textbf {\bibinfo {volume} {C77}},\ \bibinfo {pages} {688}
  (\bibinfo {year} {2017})},\ \Eprint {http://arxiv.org/abs/1704.05447}
  {arXiv:1704.05447 [hep-ph]} \BibitemShut {NoStop}%
\bibitem [{\citenamefont {Blum}\ \emph {et~al.}(2013)\citenamefont {Blum},
  \citenamefont {Denig}, \citenamefont {Logashenko}, \citenamefont {de~Rafael},
  \citenamefont {Lee~Roberts}, \citenamefont {Teubner},\ and\ \citenamefont
  {Venanzoni}}]{Blum:2013xva}%
  \BibitemOpen
  \bibfield  {author} {\bibinfo {author} {\bibfnamefont {T.}~\bibnamefont
  {Blum}}, \bibinfo {author} {\bibfnamefont {A.}~\bibnamefont {Denig}},
  \bibinfo {author} {\bibfnamefont {I.}~\bibnamefont {Logashenko}}, \bibinfo
  {author} {\bibfnamefont {E.}~\bibnamefont {de~Rafael}}, \bibinfo {author}
  {\bibfnamefont {B.}~\bibnamefont {Lee~Roberts}}, \bibinfo {author}
  {\bibfnamefont {T.}~\bibnamefont {Teubner}}, \ and\ \bibinfo {author}
  {\bibfnamefont {G.}~\bibnamefont {Venanzoni}},\ }\href@noop {} {\  (\bibinfo
  {year} {2013})},\ \Eprint {http://arxiv.org/abs/1311.2198} {arXiv:1311.2198
  [hep-ph]} \BibitemShut {NoStop}%
\bibitem [{\citenamefont {Bifani}\ \emph {et~al.}(2018)\citenamefont {Bifani},
  \citenamefont {Descotes-Genon}, \citenamefont {Romero~Vidal},\ and\
  \citenamefont {Schune}}]{Bifani:2018zmi}%
  \BibitemOpen
  \bibfield  {author} {\bibinfo {author} {\bibfnamefont {S.}~\bibnamefont
  {Bifani}}, \bibinfo {author} {\bibfnamefont {S.}~\bibnamefont
  {Descotes-Genon}}, \bibinfo {author} {\bibfnamefont {A.}~\bibnamefont
  {Romero~Vidal}}, \ and\ \bibinfo {author} {\bibfnamefont {M.-H.}\
  \bibnamefont {Schune}},\ }\href@noop {} {\  (\bibinfo {year} {2018})},\
  \Eprint {http://arxiv.org/abs/1809.06229} {arXiv:1809.06229 [hep-ex]}
  \BibitemShut {NoStop}%
\bibitem [{\citenamefont {Bobeth}\ \emph {et~al.}(2007)\citenamefont {Bobeth},
  \citenamefont {Hiller},\ and\ \citenamefont {Piranishvili}}]{Bobeth:2007dw}%
  \BibitemOpen
  \bibfield  {author} {\bibinfo {author} {\bibfnamefont {C.}~\bibnamefont
  {Bobeth}}, \bibinfo {author} {\bibfnamefont {G.}~\bibnamefont {Hiller}}, \
  and\ \bibinfo {author} {\bibfnamefont {G.}~\bibnamefont {Piranishvili}},\
  }\href {\doibase 10.1088/1126-6708/2007/12/040} {\bibfield  {journal}
  {\bibinfo  {journal} {JHEP}\ }\textbf {\bibinfo {volume} {12}},\ \bibinfo
  {pages} {040} (\bibinfo {year} {2007})},\ \Eprint
  {http://arxiv.org/abs/0709.4174} {arXiv:0709.4174 [hep-ph]} \BibitemShut
  {NoStop}%
\bibitem [{\citenamefont {Bordone}\ \emph {et~al.}(2016)\citenamefont
  {Bordone}, \citenamefont {Isidori},\ and\ \citenamefont
  {Pattori}}]{Bordone:2016gaq}%
  \BibitemOpen
  \bibfield  {author} {\bibinfo {author} {\bibfnamefont {M.}~\bibnamefont
  {Bordone}}, \bibinfo {author} {\bibfnamefont {G.}~\bibnamefont {Isidori}}, \
  and\ \bibinfo {author} {\bibfnamefont {A.}~\bibnamefont {Pattori}},\ }\href
  {\doibase 10.1140/epjc/s10052-016-4274-7} {\bibfield  {journal} {\bibinfo
  {journal} {Eur. Phys. J.}\ }\textbf {\bibinfo {volume} {C76}},\ \bibinfo
  {pages} {440} (\bibinfo {year} {2016})},\ \Eprint
  {http://arxiv.org/abs/1605.07633} {arXiv:1605.07633 [hep-ph]} \BibitemShut
  {NoStop}%
\bibitem [{\citenamefont {Glashow}\ \emph {et~al.}(2015)\citenamefont
  {Glashow}, \citenamefont {Guadagnoli},\ and\ \citenamefont
  {Lane}}]{Glashow:2014iga}%
  \BibitemOpen
  \bibfield  {author} {\bibinfo {author} {\bibfnamefont {S.~L.}\ \bibnamefont
  {Glashow}}, \bibinfo {author} {\bibfnamefont {D.}~\bibnamefont {Guadagnoli}},
  \ and\ \bibinfo {author} {\bibfnamefont {K.}~\bibnamefont {Lane}},\ }\href
  {\doibase 10.1103/PhysRevLett.114.091801} {\bibfield  {journal} {\bibinfo
  {journal} {Phys. Rev. Lett.}\ }\textbf {\bibinfo {volume} {114}},\ \bibinfo
  {pages} {091801} (\bibinfo {year} {2015})},\ \Eprint
  {http://arxiv.org/abs/1411.0565} {arXiv:1411.0565 [hep-ph]} \BibitemShut
  {NoStop}%
\bibitem [{\citenamefont {Descotes-Genon}\ \emph {et~al.}(2013)\citenamefont
  {Descotes-Genon}, \citenamefont {Matias},\ and\ \citenamefont
  {Virto}}]{Descotes-Genon:2013wba}%
  \BibitemOpen
  \bibfield  {author} {\bibinfo {author} {\bibfnamefont {S.}~\bibnamefont
  {Descotes-Genon}}, \bibinfo {author} {\bibfnamefont {J.}~\bibnamefont
  {Matias}}, \ and\ \bibinfo {author} {\bibfnamefont {J.}~\bibnamefont
  {Virto}},\ }\href {\doibase 10.1103/PhysRevD.88.074002} {\bibfield  {journal}
  {\bibinfo  {journal} {Phys. Rev.}\ }\textbf {\bibinfo {volume} {D88}},\
  \bibinfo {pages} {074002} (\bibinfo {year} {2013})},\ \Eprint
  {http://arxiv.org/abs/1307.5683} {arXiv:1307.5683 [hep-ph]} \BibitemShut
  {NoStop}%
\bibitem [{\citenamefont {Gauld}\ \emph {et~al.}(2014)\citenamefont {Gauld},
  \citenamefont {Goertz},\ and\ \citenamefont {Haisch}}]{Gauld:2013qba}%
  \BibitemOpen
  \bibfield  {author} {\bibinfo {author} {\bibfnamefont {R.}~\bibnamefont
  {Gauld}}, \bibinfo {author} {\bibfnamefont {F.}~\bibnamefont {Goertz}}, \
  and\ \bibinfo {author} {\bibfnamefont {U.}~\bibnamefont {Haisch}},\ }\href
  {\doibase 10.1103/PhysRevD.89.015005} {\bibfield  {journal} {\bibinfo
  {journal} {Phys. Rev.}\ }\textbf {\bibinfo {volume} {D89}},\ \bibinfo {pages}
  {015005} (\bibinfo {year} {2014})},\ \Eprint {http://arxiv.org/abs/1308.1959}
  {arXiv:1308.1959 [hep-ph]} \BibitemShut {NoStop}%
\bibitem [{\citenamefont {Chiang}\ \emph {et~al.}(2016)\citenamefont {Chiang},
  \citenamefont {He},\ and\ \citenamefont {Valencia}}]{Chiang:2016qov}%
  \BibitemOpen
  \bibfield  {author} {\bibinfo {author} {\bibfnamefont {C.-W.}\ \bibnamefont
  {Chiang}}, \bibinfo {author} {\bibfnamefont {X.-G.}\ \bibnamefont {He}}, \
  and\ \bibinfo {author} {\bibfnamefont {G.}~\bibnamefont {Valencia}},\ }\href
  {\doibase 10.1103/PhysRevD.93.074003} {\bibfield  {journal} {\bibinfo
  {journal} {Phys. Rev.}\ }\textbf {\bibinfo {volume} {D93}},\ \bibinfo {pages}
  {074003} (\bibinfo {year} {2016})},\ \Eprint
  {http://arxiv.org/abs/1601.07328} {arXiv:1601.07328 [hep-ph]} \BibitemShut
  {NoStop}%
\bibitem [{\citenamefont {Datta}\ \emph {et~al.}(2014)\citenamefont {Datta},
  \citenamefont {Duraisamy},\ and\ \citenamefont {Ghosh}}]{Datta:2013kja}%
  \BibitemOpen
  \bibfield  {author} {\bibinfo {author} {\bibfnamefont {A.}~\bibnamefont
  {Datta}}, \bibinfo {author} {\bibfnamefont {M.}~\bibnamefont {Duraisamy}}, \
  and\ \bibinfo {author} {\bibfnamefont {D.}~\bibnamefont {Ghosh}},\ }\href
  {\doibase 10.1103/PhysRevD.89.071501} {\bibfield  {journal} {\bibinfo
  {journal} {Phys. Rev.}\ }\textbf {\bibinfo {volume} {D89}},\ \bibinfo {pages}
  {071501} (\bibinfo {year} {2014})},\ \Eprint {http://arxiv.org/abs/1310.1937}
  {arXiv:1310.1937 [hep-ph]} \BibitemShut {NoStop}%
\bibitem [{\citenamefont {Hiller}\ and\ \citenamefont
  {Schmaltz}(2014)}]{Hiller:2014yaa}%
  \BibitemOpen
  \bibfield  {author} {\bibinfo {author} {\bibfnamefont {G.}~\bibnamefont
  {Hiller}}\ and\ \bibinfo {author} {\bibfnamefont {M.}~\bibnamefont
  {Schmaltz}},\ }\href {\doibase 10.1103/PhysRevD.90.054014} {\bibfield
  {journal} {\bibinfo  {journal} {Phys. Rev.}\ }\textbf {\bibinfo {volume}
  {D90}},\ \bibinfo {pages} {054014} (\bibinfo {year} {2014})},\ \Eprint
  {http://arxiv.org/abs/1408.1627} {arXiv:1408.1627 [hep-ph]} \BibitemShut
  {NoStop}%
\bibitem [{\citenamefont {Assad}\ \emph {et~al.}(2018)\citenamefont {Assad},
  \citenamefont {Fornal},\ and\ \citenamefont {Grinstein}}]{Assad:2017iib}%
  \BibitemOpen
  \bibfield  {author} {\bibinfo {author} {\bibfnamefont {N.}~\bibnamefont
  {Assad}}, \bibinfo {author} {\bibfnamefont {B.}~\bibnamefont {Fornal}}, \
  and\ \bibinfo {author} {\bibfnamefont {B.}~\bibnamefont {Grinstein}},\ }\href
  {\doibase 10.1016/j.physletb.2017.12.042} {\bibfield  {journal} {\bibinfo
  {journal} {Phys. Lett.}\ }\textbf {\bibinfo {volume} {B777}},\ \bibinfo
  {pages} {324} (\bibinfo {year} {2018})},\ \Eprint
  {http://arxiv.org/abs/1708.06350} {arXiv:1708.06350 [hep-ph]} \BibitemShut
  {NoStop}%
\bibitem [{\citenamefont {Di~Luzio}\ \emph {et~al.}(2017)\citenamefont
  {Di~Luzio}, \citenamefont {Greljo},\ and\ \citenamefont
  {Nardecchia}}]{DiLuzio:2017vat}%
  \BibitemOpen
  \bibfield  {author} {\bibinfo {author} {\bibfnamefont {L.}~\bibnamefont
  {Di~Luzio}}, \bibinfo {author} {\bibfnamefont {A.}~\bibnamefont {Greljo}}, \
  and\ \bibinfo {author} {\bibfnamefont {M.}~\bibnamefont {Nardecchia}},\
  }\href {\doibase 10.1103/PhysRevD.96.115011} {\bibfield  {journal} {\bibinfo
  {journal} {Phys. Rev.}\ }\textbf {\bibinfo {volume} {D96}},\ \bibinfo {pages}
  {115011} (\bibinfo {year} {2017})},\ \Eprint
  {http://arxiv.org/abs/1708.08450} {arXiv:1708.08450 [hep-ph]} \BibitemShut
  {NoStop}%
\bibitem [{\citenamefont {Calibbi}\ \emph {et~al.}(2017)\citenamefont
  {Calibbi}, \citenamefont {Crivellin},\ and\ \citenamefont
  {Li}}]{Calibbi:2017qbu}%
  \BibitemOpen
  \bibfield  {author} {\bibinfo {author} {\bibfnamefont {L.}~\bibnamefont
  {Calibbi}}, \bibinfo {author} {\bibfnamefont {A.}~\bibnamefont {Crivellin}},
  \ and\ \bibinfo {author} {\bibfnamefont {T.}~\bibnamefont {Li}},\ }\href@noop
  {} {\  (\bibinfo {year} {2017})},\ \Eprint {http://arxiv.org/abs/1709.00692}
  {arXiv:1709.00692 [hep-ph]} \BibitemShut {NoStop}%
\bibitem [{\citenamefont {Bordone}\ \emph
  {et~al.}(2018{\natexlab{a}})\citenamefont {Bordone}, \citenamefont
  {Cornella}, \citenamefont {Fuentes-Martin},\ and\ \citenamefont
  {Isidori}}]{Bordone:2017bld}%
  \BibitemOpen
  \bibfield  {author} {\bibinfo {author} {\bibfnamefont {M.}~\bibnamefont
  {Bordone}}, \bibinfo {author} {\bibfnamefont {C.}~\bibnamefont {Cornella}},
  \bibinfo {author} {\bibfnamefont {J.}~\bibnamefont {Fuentes-Martin}}, \ and\
  \bibinfo {author} {\bibfnamefont {G.}~\bibnamefont {Isidori}},\ }\href
  {\doibase 10.1016/j.physletb.2018.02.011} {\bibfield  {journal} {\bibinfo
  {journal} {Phys. Lett.}\ }\textbf {\bibinfo {volume} {B779}},\ \bibinfo
  {pages} {317} (\bibinfo {year} {2018}{\natexlab{a}})},\ \Eprint
  {http://arxiv.org/abs/1712.01368} {arXiv:1712.01368 [hep-ph]} \BibitemShut
  {NoStop}%
\bibitem [{\citenamefont {Barbieri}\ and\ \citenamefont
  {Tesi}(2018)}]{Barbieri:2017tuq}%
  \BibitemOpen
  \bibfield  {author} {\bibinfo {author} {\bibfnamefont {R.}~\bibnamefont
  {Barbieri}}\ and\ \bibinfo {author} {\bibfnamefont {A.}~\bibnamefont
  {Tesi}},\ }\href {\doibase 10.1140/epjc/s10052-018-5680-9} {\bibfield
  {journal} {\bibinfo  {journal} {Eur. Phys. J.}\ }\textbf {\bibinfo {volume}
  {C78}},\ \bibinfo {pages} {193} (\bibinfo {year} {2018})},\ \Eprint
  {http://arxiv.org/abs/1712.06844} {arXiv:1712.06844 [hep-ph]} \BibitemShut
  {NoStop}%
\bibitem [{\citenamefont {Blanke}\ and\ \citenamefont
  {Crivellin}(2018)}]{Blanke:2018sro}%
  \BibitemOpen
  \bibfield  {author} {\bibinfo {author} {\bibfnamefont {M.}~\bibnamefont
  {Blanke}}\ and\ \bibinfo {author} {\bibfnamefont {A.}~\bibnamefont
  {Crivellin}},\ }\href {\doibase 10.1103/PhysRevLett.121.011801} {\bibfield
  {journal} {\bibinfo  {journal} {Phys. Rev. Lett.}\ }\textbf {\bibinfo
  {volume} {121}},\ \bibinfo {pages} {011801} (\bibinfo {year} {2018})},\
  \Eprint {http://arxiv.org/abs/1801.07256} {arXiv:1801.07256 [hep-ph]}
  \BibitemShut {NoStop}%
\bibitem [{\citenamefont {Greljo}\ and\ \citenamefont
  {Stefanek}(2018)}]{Greljo:2018tuh}%
  \BibitemOpen
  \bibfield  {author} {\bibinfo {author} {\bibfnamefont {A.}~\bibnamefont
  {Greljo}}\ and\ \bibinfo {author} {\bibfnamefont {B.~A.}\ \bibnamefont
  {Stefanek}},\ }\href {\doibase 10.1016/j.physletb.2018.05.033} {\bibfield
  {journal} {\bibinfo  {journal} {Phys. Lett.}\ }\textbf {\bibinfo {volume}
  {B782}},\ \bibinfo {pages} {131} (\bibinfo {year} {2018})},\ \Eprint
  {http://arxiv.org/abs/1802.04274} {arXiv:1802.04274 [hep-ph]} \BibitemShut
  {NoStop}%
\bibitem [{\citenamefont {Bordone}\ \emph
  {et~al.}(2018{\natexlab{b}})\citenamefont {Bordone}, \citenamefont
  {Cornella}, \citenamefont {Fuentes-Martín},\ and\ \citenamefont
  {Isidori}}]{Bordone:2018nbg}%
  \BibitemOpen
  \bibfield  {author} {\bibinfo {author} {\bibfnamefont {M.}~\bibnamefont
  {Bordone}}, \bibinfo {author} {\bibfnamefont {C.}~\bibnamefont {Cornella}},
  \bibinfo {author} {\bibfnamefont {J.}~\bibnamefont {Fuentes-Martín}}, \ and\
  \bibinfo {author} {\bibfnamefont {G.}~\bibnamefont {Isidori}},\ }\href@noop
  {} {\  (\bibinfo {year} {2018}{\natexlab{b}})},\ \Eprint
  {http://arxiv.org/abs/1805.09328} {arXiv:1805.09328 [hep-ph]} \BibitemShut
  {NoStop}%
\bibitem [{\citenamefont {Faber}\ \emph {et~al.}(2018)\citenamefont {Faber},
  \citenamefont {Hudec}, \citenamefont {Malinský}, \citenamefont {Meinzinger},
  \citenamefont {Porod},\ and\ \citenamefont {Staub}}]{Faber:2018qon}%
  \BibitemOpen
  \bibfield  {author} {\bibinfo {author} {\bibfnamefont {T.}~\bibnamefont
  {Faber}}, \bibinfo {author} {\bibfnamefont {M.}~\bibnamefont {Hudec}},
  \bibinfo {author} {\bibfnamefont {M.}~\bibnamefont {Malinský}}, \bibinfo
  {author} {\bibfnamefont {P.}~\bibnamefont {Meinzinger}}, \bibinfo {author}
  {\bibfnamefont {W.}~\bibnamefont {Porod}}, \ and\ \bibinfo {author}
  {\bibfnamefont {F.}~\bibnamefont {Staub}},\ }\href@noop {} {\  (\bibinfo
  {year} {2018})},\ \Eprint {http://arxiv.org/abs/1808.05511} {arXiv:1808.05511
  [hep-ph]} \BibitemShut {NoStop}%
\bibitem [{\citenamefont {Heeck}\ and\ \citenamefont
  {Teresi}(2018)}]{Heeck:2018ntp}%
  \BibitemOpen
  \bibfield  {author} {\bibinfo {author} {\bibfnamefont {J.}~\bibnamefont
  {Heeck}}\ and\ \bibinfo {author} {\bibfnamefont {D.}~\bibnamefont {Teresi}},\
  }\href@noop {} {\  (\bibinfo {year} {2018})},\ \Eprint
  {http://arxiv.org/abs/1808.07492} {arXiv:1808.07492 [hep-ph]} \BibitemShut
  {NoStop}%
\bibitem [{\citenamefont {Pati}\ and\ \citenamefont
  {Salam}(1974)}]{Pati:1974yy}%
  \BibitemOpen
  \bibfield  {author} {\bibinfo {author} {\bibfnamefont {J.~C.}\ \bibnamefont
  {Pati}}\ and\ \bibinfo {author} {\bibfnamefont {A.}~\bibnamefont {Salam}},\
  }\href {\doibase 10.1103/PhysRevD.10.275, 10.1103/PhysRevD.11.703.2}
  {\bibfield  {journal} {\bibinfo  {journal} {Phys. Rev.}\ }\textbf {\bibinfo
  {volume} {D10}},\ \bibinfo {pages} {275} (\bibinfo {year} {1974})},\ \bibinfo
  {note} {[Erratum: Phys. Rev.D11,703(1975)]}\BibitemShut {NoStop}%
\bibitem [{\citenamefont {Foot}(1998)}]{Foot:1997pb}%
  \BibitemOpen
  \bibfield  {author} {\bibinfo {author} {\bibfnamefont {R.}~\bibnamefont
  {Foot}},\ }\href {\doibase 10.1016/S0370-2693(97)01519-0} {\bibfield
  {journal} {\bibinfo  {journal} {Phys. Lett.}\ }\textbf {\bibinfo {volume}
  {B420}},\ \bibinfo {pages} {333} (\bibinfo {year} {1998})},\ \Eprint
  {http://arxiv.org/abs/hep-ph/9708205} {arXiv:hep-ph/9708205 [hep-ph]}
  \BibitemShut {NoStop}%
\bibitem [{\citenamefont {Foot}\ and\ \citenamefont
  {Filewood}(1999)}]{Foot:1999wv}%
  \BibitemOpen
  \bibfield  {author} {\bibinfo {author} {\bibfnamefont {R.}~\bibnamefont
  {Foot}}\ and\ \bibinfo {author} {\bibfnamefont {G.}~\bibnamefont
  {Filewood}},\ }\href {\doibase 10.1103/PhysRevD.60.115002} {\bibfield
  {journal} {\bibinfo  {journal} {Phys. Rev.}\ }\textbf {\bibinfo {volume}
  {D60}},\ \bibinfo {pages} {115002} (\bibinfo {year} {1999})},\ \Eprint
  {http://arxiv.org/abs/hep-ph/9903374} {arXiv:hep-ph/9903374 [hep-ph]}
  \BibitemShut {NoStop}%
\bibitem [{\citenamefont {Haber}\ and\ \citenamefont
  {Nir}(1990)}]{Haber:1989xc}%
  \BibitemOpen
  \bibfield  {author} {\bibinfo {author} {\bibfnamefont {H.~E.}\ \bibnamefont
  {Haber}}\ and\ \bibinfo {author} {\bibfnamefont {Y.}~\bibnamefont {Nir}},\
  }\href {\doibase 10.1016/0550-3213(90)90499-4} {\bibfield  {journal}
  {\bibinfo  {journal} {Nucl. Phys.}\ }\textbf {\bibinfo {volume} {B335}},\
  \bibinfo {pages} {363} (\bibinfo {year} {1990})}\BibitemShut {NoStop}%
\bibitem [{\citenamefont {Aad}\ \emph {et~al.}(2012)\citenamefont {Aad} \emph
  {et~al.}}]{Aad:2012tfa}%
  \BibitemOpen
  \bibfield  {author} {\bibinfo {author} {\bibfnamefont {G.}~\bibnamefont
  {Aad}} \emph {et~al.} (\bibinfo {collaboration} {ATLAS}),\ }\href {\doibase
  10.1016/j.physletb.2012.08.020} {\bibfield  {journal} {\bibinfo  {journal}
  {Phys. Lett.}\ }\textbf {\bibinfo {volume} {B716}},\ \bibinfo {pages} {1}
  (\bibinfo {year} {2012})},\ \Eprint {http://arxiv.org/abs/1207.7214}
  {arXiv:1207.7214 [hep-ex]} \BibitemShut {NoStop}%
\bibitem [{\citenamefont {Chatrchyan}\ \emph {et~al.}(2012)\citenamefont
  {Chatrchyan} \emph {et~al.}}]{Chatrchyan:2012xdj}%
  \BibitemOpen
  \bibfield  {author} {\bibinfo {author} {\bibfnamefont {S.}~\bibnamefont
  {Chatrchyan}} \emph {et~al.} (\bibinfo {collaboration} {CMS}),\ }\href
  {\doibase 10.1016/j.physletb.2012.08.021} {\bibfield  {journal} {\bibinfo
  {journal} {Phys. Lett.}\ }\textbf {\bibinfo {volume} {B716}},\ \bibinfo
  {pages} {30} (\bibinfo {year} {2012})},\ \Eprint
  {http://arxiv.org/abs/1207.7235} {arXiv:1207.7235 [hep-ex]} \BibitemShut
  {NoStop}%
\bibitem [{\citenamefont {Straub}\ \emph {et~al.}(2018)\citenamefont {Straub},
  \citenamefont {Stangl}, \citenamefont {Niehoff}, \citenamefont {G\"urler},
  \citenamefont {Wang}, \citenamefont {Kumar}, \citenamefont {Reichert},\ and\
  \citenamefont {Beaujean}}]{david_straub_2018_1326349}%
  \BibitemOpen
  \bibfield  {author} {\bibinfo {author} {\bibfnamefont {D.}~\bibnamefont
  {Straub}}, \bibinfo {author} {\bibfnamefont {P.}~\bibnamefont {Stangl}},
  \bibinfo {author} {\bibfnamefont {C.}~\bibnamefont {Niehoff}}, \bibinfo
  {author} {\bibfnamefont {E.}~\bibnamefont {G\"urler}}, \bibinfo {author}
  {\bibfnamefont {Z.}~\bibnamefont {Wang}}, \bibinfo {author} {\bibfnamefont
  {J.}~\bibnamefont {Kumar}}, \bibinfo {author} {\bibfnamefont
  {S.}~\bibnamefont {Reichert}}, \ and\ \bibinfo {author} {\bibfnamefont
  {F.}~\bibnamefont {Beaujean}},\ }\href {\doibase 10.5281/zenodo.1326349}
  {\enquote {\bibinfo {title} {flav-io/flavio v0.29.1},}\ } (\bibinfo {year}
  {2018})\BibitemShut {NoStop}%
\bibitem [{\citenamefont {Tanabashi}\ \emph {et~al.}(2018)\citenamefont
  {Tanabashi} \emph {et~al.}}]{Tanabashi:2018oca}%
  \BibitemOpen
  \bibfield  {author} {\bibinfo {author} {\bibfnamefont {M.}~\bibnamefont
  {Tanabashi}} \emph {et~al.} (\bibinfo {collaboration} {ParticleDataGroup}),\
  }\href {\doibase 10.1103/PhysRevD.98.030001} {\bibfield  {journal} {\bibinfo
  {journal} {Phys. Rev.}\ }\textbf {\bibinfo {volume} {D98}},\ \bibinfo {pages}
  {030001} (\bibinfo {year} {2018})}\BibitemShut {NoStop}%
\bibitem [{\citenamefont {Aaij}\ \emph {et~al.}(2018)\citenamefont {Aaij} \emph
  {et~al.}}]{Bediaga:2018lhg}%
  \BibitemOpen
  \bibfield  {author} {\bibinfo {author} {\bibfnamefont {R.}~\bibnamefont
  {Aaij}} \emph {et~al.} (\bibinfo {collaboration} {LHCb}),\ }\href@noop {} {\
  (\bibinfo {year} {2018})},\ \Eprint {http://arxiv.org/abs/1808.08865}
  {arXiv:1808.08865} \BibitemShut {NoStop}%
\bibitem [{\citenamefont {Kou}\ \emph {et~al.}(2018)\citenamefont {Kou} \emph
  {et~al.}}]{Kou:2018nap}%
  \BibitemOpen
  \bibfield  {author} {\bibinfo {author} {\bibfnamefont {E.}~\bibnamefont
  {Kou}} \emph {et~al.} (\bibinfo {collaboration} {Belle II}),\ }\href@noop {}
  {\  (\bibinfo {year} {2018})},\ \Eprint {http://arxiv.org/abs/1808.10567}
  {arXiv:1808.10567 [hep-ex]} \BibitemShut {NoStop}%
\bibitem [{\citenamefont {Bobeth}\ \emph {et~al.}(2014)\citenamefont {Bobeth},
  \citenamefont {Gorbahn}, \citenamefont {Hermann}, \citenamefont {Misiak},
  \citenamefont {Stamou},\ and\ \citenamefont {Steinhauser}}]{Bobeth:2013uxa}%
  \BibitemOpen
  \bibfield  {author} {\bibinfo {author} {\bibfnamefont {C.}~\bibnamefont
  {Bobeth}}, \bibinfo {author} {\bibfnamefont {M.}~\bibnamefont {Gorbahn}},
  \bibinfo {author} {\bibfnamefont {T.}~\bibnamefont {Hermann}}, \bibinfo
  {author} {\bibfnamefont {M.}~\bibnamefont {Misiak}}, \bibinfo {author}
  {\bibfnamefont {E.}~\bibnamefont {Stamou}}, \ and\ \bibinfo {author}
  {\bibfnamefont {M.}~\bibnamefont {Steinhauser}},\ }\href {\doibase
  10.1103/PhysRevLett.112.101801} {\bibfield  {journal} {\bibinfo  {journal}
  {Phys. Rev. Lett.}\ }\textbf {\bibinfo {volume} {112}},\ \bibinfo {pages}
  {101801} (\bibinfo {year} {2014})},\ \Eprint {http://arxiv.org/abs/1311.0903}
  {arXiv:1311.0903 [hep-ph]} \BibitemShut {NoStop}%
\bibitem [{\citenamefont {Glattauer}\ \emph {et~al.}(2016)\citenamefont
  {Glattauer} \emph {et~al.}}]{Glattauer:2015teq}%
  \BibitemOpen
  \bibfield  {author} {\bibinfo {author} {\bibfnamefont {R.}~\bibnamefont
  {Glattauer}} \emph {et~al.} (\bibinfo {collaboration} {Belle}),\ }\href
  {\doibase 10.1103/PhysRevD.93.032006} {\bibfield  {journal} {\bibinfo
  {journal} {Phys. Rev.}\ }\textbf {\bibinfo {volume} {D93}},\ \bibinfo {pages}
  {032006} (\bibinfo {year} {2016})},\ \Eprint
  {http://arxiv.org/abs/1510.03657} {arXiv:1510.03657 [hep-ex]} \BibitemShut
  {NoStop}%
\bibitem [{\citenamefont {Abdesselam}\ \emph {et~al.}(2017)\citenamefont
  {Abdesselam} \emph {et~al.}}]{Abdesselam:2017kjf}%
  \BibitemOpen
  \bibfield  {author} {\bibinfo {author} {\bibfnamefont {A.}~\bibnamefont
  {Abdesselam}} \emph {et~al.} (\bibinfo {collaboration} {Belle}),\ }\href@noop
  {} {\  (\bibinfo {year} {2017})},\ \Eprint {http://arxiv.org/abs/1702.01521}
  {arXiv:1702.01521 [hep-ex]} \BibitemShut {NoStop}%
\bibitem [{\citenamefont {Lavoura}(2003)}]{Lavoura:2003xp}%
  \BibitemOpen
  \bibfield  {author} {\bibinfo {author} {\bibfnamefont {L.}~\bibnamefont
  {Lavoura}},\ }\href {\doibase 10.1140/epjc/s2003-01212-7} {\bibfield
  {journal} {\bibinfo  {journal} {Eur. Phys. J.}\ }\textbf {\bibinfo {volume}
  {C29}},\ \bibinfo {pages} {191} (\bibinfo {year} {2003})},\ \Eprint
  {http://arxiv.org/abs/hep-ph/0302221} {arXiv:hep-ph/0302221 [hep-ph]}
  \BibitemShut {NoStop}%
\end{thebibliography}%

\end{document}